\documentclass[useAMS,usenatbib]{mn2e}
\usepackage{float,multicol,epsfig}

\title[Spectral distortions to the CMB]
{Spectral distortions to the Cosmic Microwave Background from
the recombination of hydrogen and helium}

\author[W. Y. Wong, S. Seager and D. Scott]{Wan Yan Wong$^{1}$\thanks{E-mail:
wanyan@phas.ubc.ca},
 Sara Seager$^{2}$\thanks{E-mail: seager@dtm.ciw.edu}
and Douglas Scott$^{1}$\thanks{E-mail: dscott@phas.ubc.ca}
\\
$^{1}$Department of Physics and Astronomy, University of British Columbia,
 6224 Agricultural Rd., Vancouver, BC, V6T 1Z1, Canada\\
$^{2}$Department of Terrestrial Magnetism, Carnegie Institution of Washington,
 5241 Broad Branch Rd. NW, Washington, DC 20015, USA}

\begin{document}
\date{2005 October 19}

\pagerange{\pageref{firstpage}--\pageref{lastpage}} \pubyear{2005}

\maketitle

\label{firstpage}
\begin{abstract}
The recombination of hydrogen and helium at $z\,{\sim}\,1000$--7000 gives
unavoidable distortions to the Cosmic Microwave Background (CMB) spectrum.
We present a detailed calculation of the line intensities arising from
the Ly$\,\alpha$ (2p--1s) and two-photon (2s--1s) transitions for the
recombination of hydrogen, as well as the corresponding lines from
helium.  We give an approximate formula for the strength of the main
recombination line distortion on the CMB in different cosmologies, this
peak occurring at about $170\,\mu$m.
We also find a previously undescribed long wavelength peak (which we call
the pre-recombination peak) from the lines of the 2p--1s transitions,
which are formed before significant recombination of the corresponding atoms
occurred.
Detailed calculations of the two-photon emission line shapes are presented
here for the first time.  The frequencies of the photons emitted from
the two-photon transition have a wide spectrum and this causes the
location of the peak of the two-photon line of hydrogen to be located almost
at the same wavelength as the main Ly$\,\alpha$ peak.  The helium lines also
give distortions at similar wavelengths, so that the combined distortion has
a complex shape.
The detection of this distortion would
provide direct supporting evidence that the Universe was indeed once a
plasma.  Moreover, the distortions are a sensitive probe of physics
during the time of recombination.  Although the spectral
distortion is overwhelmed by dust emission from the Galaxy, and is maximum
at wavelengths roughly where the cosmic far-infrared background peaks,
it may be able to tailor an experiment to detect its non-trivial shape.
\end{abstract}

\section{Introduction}
Physical processes in the plasma of the hot early Universe thermalize
the radiation content, and this redshifts to become the observed
Cosmic Microwave Background \citep[CMB; see][and references
therein]{scott04}.  Besides the photons from the radiation background,
there were some extra photons produced from the transitions when the
electrons cascaded down to the ground state after they recombined with
the ionized atoms.  The transition from a plasma to mainly neutral gas
occurred because as the Universe expanded the background temperature
dropped, allowing the ions to hold onto their electrons.  The photons
created in this process give a distortion to the nearly perfect
blackbody CMB spectrum.  Since recombination happens at redshift
$z\,{\sim}\,1000$, then Ly$\,\alpha$ is observed at ${\sim}\,100\mu$m
today.  There is approximately one of these photons per baryon, which
should be compared with the ${\sim}\,10^9$ photons per baryon in the
entire CMB.  However, the recombination photons are superimposed on
the Wien part of the CMB spectrum, and so make a potentially
measurable distortion.

From the Far-Infrared Absolute Spectrophotometer (FIRAS) measurements,
\citet{fixsen96} and \citet{mather99} showed that the CMB is well
modelled by a $2.725\pm0.001\,$K blackbody, and that any
deviations from this spectrum around the peak are less than 50 parts per
million of the peak brightness.  Constraints on smooth functions, such as
$\mu$- or $y$-distortions are similarly very stringent.
However, there are much weaker constraints on narrower features in the CMB
spectrum.  Moreover, within the last decade it has been discovered
\cite{puget96} that there is a Cosmic Infrared Background~\citep[CIB;
see][and references therein]{hauserdwek01}, which peaks at $100$--$200\mu$m
and is mainly composed of luminous infrared galaxies at moderate redshifts.
The existence of this background makes it more challenging to measure the
recombination distortions than would have been the case if one imagined them
only as being distortions to Wien tail of the CMB.  However, as we shall see,
the shape of the recombination line distortion is expected to be much narrower
than that of the CIB, and hence the signal may be detectable in a future
experiment designed to measure the CIB spectrum in detail.

The first published calculations of the line distortions occur in the
seminal papers on the cosmological recombination process by
\citet{peebles1968} and Zel'dovich, Kurt \& Sunyaev (1968).  One of
the main motivations for studying the recombination process was to
answer the question: `Where are the Ly$\,\alpha$ line photons from the
recombination in the Universe?'~(as reported in Rubino-Martin, 
Hernandez-Monteagudo \& Sunyaev 2005).
In fact these studies found that for hydrogen recombination (in a
cosmology which is somewhat different than the model favoured today)
there are more photons created through the two-photon 2s--1s
transition than from the Ly$\,\alpha$ transition.  Both
\citet{peebles1968} and \citet{zks68} plot the distortion of the CMB
tail caused by these line photons, but give no detail about the line
shapes.  Other authors have included some calculation or discussion of
the line distortions as part of other recombination related studies,
e.g.~\citet{boschan98}, and most recently \citet{switzer05}.  However,
the explicit line shapes have never before been presented, and the
helium lines have also been neglected so far.  The only numerical
study to show the hydrogen lines in any detail is a short conference
report by \citet{dell93}, meant as a preliminary version of a more
full study which never appeared.  Although their calculation appears
to have been substantially correct, unfortunately in the one plot they
show of the distortions (their fig.~2) it is difficult to tell
precisely which effects are real and which might be numerical.

Some of the recombination line distortions
from higher energy levels, $n>2$, have also been
calculated~\citep{dub75, lyu83, fahr91, bur94, dell93, dubs95,
dubs97, burgin03,kho05}.  However, these high $n$ lines are extremely weak
compared with the CMB (below the $10^{-6}$ level), while the
Ly$\,\alpha$ line is well above the CMB in the Wien region of the spectrum.

As trumpeted by many authors, we are now entering into the era of
precision cosmology.  Hence one might imagine that future delicate
experiments may be able to measure these line distortions.
Since the lines are formed by the photons emitted in each transitions of
the electrons, they are strongly dependent on the rate of recombination
of the atoms.  The distortion lines may thus be a more sensitive probe of
recombination era physics than the ionization fraction
$x_{\rm e}$, and the related visibility function which affects the CMB
anisotropies.  This is because a lot of energy must be injected in order
for any physical process to change
$x_{\rm e}$ substantially~\citep[e.g.][]{psh00}.  In general that energy will
go into spectral distortions, including boosting the recombination lines.

This also means that a detailed understanding of the
physics of recombination is crucial for calculating the
distortion.  The basic physical picture for cosmological recombination
has not changed since the early work of \citet{peebles1968} and \citet{zks68}.
However, there have been several refinements introduced since then, motivated
by the increased emphasis on obtaining an accurate recombination
history as part of the calculation of CMB anisotropies.
Seager, Sasselov \& Scott (1999,2000) presented a detailed
calculation of the whole recombination process, with no assumption of
equilibrium among the energy levels.  This multi-level computation
involves 300 levels for both hydrogen and helium, and gives us the
currently most accurate picture of the recombination history.  In the
context of the \citet{sara00} recombination calculation, and with the
well-developed set of cosmological parameters provided by Wilkinson
Microwave Anisotropy Probe~\citep[{\sl WMAP\/};][]{spergel03} and other
CMB experiments, it seems an appropriate time to calculate the distortion
lines to higher accuracy in order to investigate whether they could be
detected and whether their detection might be cosmologically useful.

The aim of this paper is to calculate the line distortions on the CMB
from the 2p--1s and 2s--1s transitions of H and the corresponding
lines of He~(i.e. the $2^1$p--$1^1$s and $2^1$s--$1^1$s transitions of
He{\small I}, and the 2p--1s and 2s--1s transitions of He{\small II})
during recombination, using the standard cosmological parameters and
recombination history.  In Section~\ref{theory} we will describe the
model we used in the numerical calculation and give the equations used to
calculate the spectral lines.  In Section~\ref{result} we will present our
results and discuss the detailed physics of the locations and shapes of
the spectral lines.  An approximate formula for the magnitude of the
distortion in different cosmologies will also be given.  Other
possible modifications of the spectral lines and their potential
detectability will be discussed in
Section~\ref{discuss}.  And finally, we will present our conclusions
in the last section.

\section{Basic theory}
\label{theory}
\subsection{Model}
Instead of adopting a full multi-level code, we use a simple 3-level
model atom here.  For single-electron atoms (i.e. H{\small I} and
He{\small II}), we consider only the ground state, the first excited
state and the continuum.  For the 2-electron atom (He{\small I}), we
consider the corresponding levels among singlet states.  In general,
the upper level states are considered to be in thermal equilibrium
with the first excited state.  Case B recombination is adopted here,
which means that we ignore recombinations and photo-ionizations
directly to ground state.  This is because the photons emitted from
direct recombinations to the ground state will almost
immediately reionize a nearby neutral H atom \citep{peebles1968,
sara00}.  We also include the two-photon rate from 2s to the ground
state for all three atoms, with rates: $\Lambda^{\rm H}_{\rm 2s-1s} =
8.229063\,$s$^{-1}$ \citep{gold89,santos98}; $\Lambda^{\rm HeI}_{\rm 2^1s-1^1s}
= 51.02\,$s$^{-1}$\citep{derev97}, although it makes no noticeable
difference to the calculation if one uses the older value of
$51.3\,$s$^{-1}$ from Drake, Victor \& Dalgamrno~(1969);
and $\Lambda^{\rm HeII}_{\rm 2s-1s} = 526.532\,$s$^{-1}$ \citep{lip65, gold89}.
This 3-level atom model is similar to the one used in the program {\small
recfast}, with the main difference being that here we
do not assume that the rate of change of the first
excited state $n_2$ is zero.

The rate equations for the 3 atoms are
similar, and so we will just state the hydrogen case as an example:
{\setlength\arraycolsep{2pt}
\begin{eqnarray}
(1+z) \frac{dn_1^{\rm{H}}(z)}{dz} &=& -\frac{1}{H(z)}
[\Delta R_{2\mathrm{p}-1\mathrm{s}}^{\rm{H}}
+ \Delta R_{2\mathrm{s}-1\mathrm{s}}^{\rm{H}} ] +3n_1^{\rm{H}} ; \\
(1+z) \frac{dn_2^{\rm{H}}(z)}{dz} &=& -\frac{1}{H(z)} [
n_{\rm{e}}n_{\rm{p}} \alpha_{\rm{H}}
- n_{2\rm{s}}^{\rm{H}} \beta_{\rm{H}} \nonumber  \\
&& \quad - \Delta R_{2\mathrm{p}-1\mathrm{s}}^{\rm{H}}
- \Delta R_{2\mathrm{s}-1\mathrm{s}}^{\rm{H}} ] +3n_2^{\rm{H}} ; \\
(1+z) \frac{dn_{\rm{e}}(z)}{dz} &=& -\frac{1}{H(z)} \left[
n_{2\rm{s}}^{\rm{H}} \beta_{\rm{H}} - n_{\rm{e}}n_{\rm{p}} \alpha_{\rm{H}}
 \right] +3n_{\rm{e}} ; \\
(1+z) \frac{dn_{\rm{p}}(z)}{dz} &=& -\frac{1}{H(z)} \left[
n_{2\rm{s}}^{\rm{H}} \beta_{\rm{H}} - n_{\rm{e}}n_{\rm{p}} \alpha_{\rm{H}}
 \right] +3n_{\rm{p}}.
\end{eqnarray}}

\noindent Here the values of $n_i$ are the number density of the $i$th
state, where $n_{\rm e}$ and $n_{\rm p}$ are the number density of
electrons and protons respectively. $\Delta R^{\rm H}_{i-j}$ is the
net bound-bound rate between state $i$ and $j$ and the detailed form of
$\Delta R_{2\mathrm{p}-1\mathrm{s}}^{\rm{H}}$ and $ \Delta
R_{2\mathrm{s}-1\mathrm{s}}^{\rm{H}}$ will be discussed in the
next subsection.  $H(z) \equiv \dot{a}/a$ is the Hubble factor,
{\setlength\arraycolsep{2pt}
\begin{eqnarray}
H(z)^2 &= & H_0^2 \bigg[ \frac{\Omega_{\rm M}}{1+z_{\rm eq}}(1+z)^4 \nonumber \\
&& + \Omega_{\rm M} (1+z)^3 + \Omega_K(1+z)^2 + \Omega_{\Lambda} \bigg],
\end{eqnarray}}

\noindent where $\Omega$ represents the fraction of the critical
density in matter, curvature or cosmological constant, and the Hubble
parameter today $H_0=100h$~km~s$^{-1}$Mpc$^{-1}$.  Finally $\alpha_{\rm{H}}$ is
the Case B recombination coefficient from \citet{hummer94},
\begin{equation}
\alpha_{\rm{H}} =  \ 10^{-19} \frac{at^b}{1+ct^d} \ \rm{m}^3 s^{-1},
\end{equation}
which is fitted by \citet{pequignot91}, with $a=4.309$, $b=-0.6166$,
$c=0.6703$, $d=0.5300$ and $t=T_{\rm{M}}/10^4$K, while $\beta_{\rm{H}}$ is the
photo-ionization coefficient:
\begin{equation}
\beta_{\rm{H}} = \alpha_{\rm{H}} \left( \frac{2 \pi m_{\rm e}
k_{\rm{B}} T_{\rm{M}}}
{h_{\rm p}^2} \right)^{\frac{3}{2}} {\rm exp} \left\{
-\frac{h_{\rm p} \nu_{2\rm{s-c}}}{k_{\rm{B}} T_{\rm{M}}} \right\},
\label{betaH}
\end{equation}
where $k_{\rm B}$ is Boltzmann's constant, $h_{\rm p}$ is Planck's
constant, $m_{\rm e}$ is the mass of electron, $T_{\rm M}$ is the matter
temperature and $\nu_{\rm 2s-c}$ is the frequency of the energy difference
between state 2s and the continuum.  For the rate of change of $T_{\rm M}$,
we include only the Compton and adiabatic cooling terms \citep{sara00}, i.e.
\begin{equation}
(1+z) \frac{d T_{\rm{M}}} {dz} =
\frac{8 \sigma_{\rm{T}} U}{3 H(z) m_e c}
\frac{n_e}{n_e + n_{\rm{H}} + n_{\rm{He}} } (T_{\rm{M}}-T_{\rm{R}})
+ 2 T_{\rm{M}},
\label{eqTM}
\end{equation}
where $T_{\rm R}$ is the radiation temperature, $c$ is the
speed of light,  $U= a_{\mathrm{R}}
T_{\mathrm{R}}^4$, $a_{\rm R}$ is the radiation constant and
$\sigma_{\mathrm{T}}$ is the Thompson scattering cross-section.

We use the Bader-Deuflhard semi-implicit numerical integration scheme
~\citep[see Section 16.6 in][]{nr} to solve the above rate equations.
All the numerical results are made using the $\Lambda$CDM model with
parameters: $\Omega_{\rm B}=0.046$; $\Omega_{\rm M}=0.3$;
$\Omega_{\Lambda}=0.7$; $\Omega_{\rm K}=0$; $Y_p=0.24$; $T_0=2.725\,$K
and $h=0.7$~\citep[see e.g.][]{spergel03}. Here $Y_{\mathrm{p}}$ is
the primordial He abundance and $T_0$ the present background
temperature.
\subsection{Spectral distortions}
\begin{figure}
\centering
\vspace*{6cm}
\leavevmode
\includegraphics{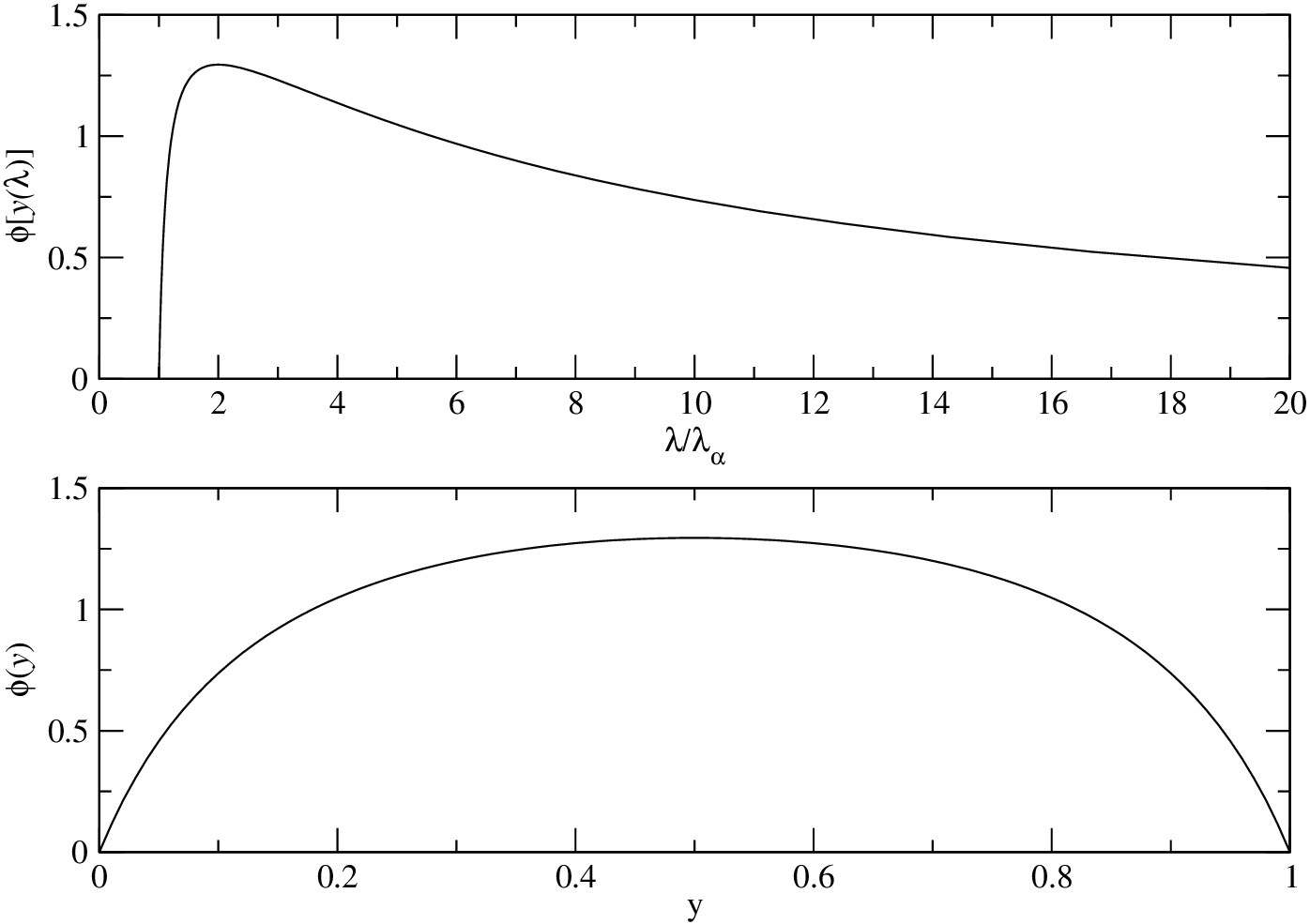}
\caption{\label{phi_H} The normalized emission spectrum for
the two-photon process (2s--1s) of hydrogen~\citep{spitzer51,
martinis00}.  The top panel shows $\phi[y(\lambda)]$ vs $\lambda$, while
the bottom panel shows $\phi(y)$ vs $y$, where $\nu = y \nu_\alpha$.
Note that the spectrum is symmetric in $\nu$ about $\nu_{\alpha}/2$,
but the $\lambda$ spectrum is very asymmetric, being zero below
$\lambda_{\alpha}$, and having a tail extending to high $\lambda$.}
\end{figure}

We want to calculate the specific line intensity $I_{\nu_0}(z=0)$
(i.e. energy per unit time per unit area per unit frequency per unit
solid angle, measured in W$\,$m$^{-2}$Hz$^{-1}{\rm sr}^{-1}$) observed
at the present epoch, $z=0$.  The detailed calculation of
$I_{\nu_0}(z=0)$ for the Ly$\,\alpha$ transition and the two-photon
transition in hydrogen are presented as examples~\citep[the notation
follows Section 2.5 in][]{pad93}.  A similar derivation holds for the
corresponding transitions in helium.  To perform these calculations we
first consider the emissivity $j_{\nu}(z)$ (energy per unit time per
unit volume per unit frequency, measured in W$\,$m$^{-3}$Hz$^{-1}$) of
photons due to the transition of electrons between the 2p and 1s
states at redshift $z$:
\begin{equation}
j_{\nu}(z)=h_{\mathrm{p}} \nu \Delta R_{2\rm{p}-1\rm{s}}^{\rm{H}}(z)
\phi[\nu(z)],
\end{equation}
where $\phi(\nu)$ is the frequency distribution of the emitted photons
from the emission process and $\Delta R_{2\rm{p}-1\rm{s}}^{\rm{H}}$ is
the net rate of photon production between the 2p and 1s levels, i.e.
\begin{equation}
\Delta R_{2\rm{p}-1\rm{s}}^{\rm{H}} = p_{12} \left(
n_{2 \rm{p}}^{\rm{H}} R_{21} - n_1^{\rm{H}} R_{12} \right).
\label{RLyH}
\end{equation}
Here $n_i^{\rm{H}}$ is the number density of hydrogen atoms having
electrons in state $i$, the upward and downward transition rates are
\begin{eqnarray}
R_{12} &=&  B_{12} \bar{J}, \\
\mathrm{and} \quad R_{21} &=& \left(A_{21} + B_{21} \bar{J}\right),
\end{eqnarray}
with $A_{21}$, $B_{12}$ and $ B_{21}$ being the Einstein coefficients
and $p_{12}$ the Sobolev escape probability~\citep[see][]{sara00},
which accounts for the redshifting of the Ly$\,\alpha$ photons due to
the expansion of the Universe.  As $n^{\rm H}_{1} \gg n^{\rm H}_{2
\rm{p}} $, $p_{12}$ can be expressed in the following form:
\begin{equation}
p_{12} = \frac{1- e^{-\tau_{\rm s}}}{\tau_{\rm s}}, \rm{with}
\label{es_prob}
\end{equation}
\begin{equation}
\tau_{\rm s}= \frac{A_{21} \lambda_{\rm 2p-1s}^3 \left(g_{2\rm{p}}
/g_1 \right) n_1}
{8 \pi H(z)}.
\end{equation}
We approximate the background radiation field $\bar{J}$ as a perfect
blackbody spectrum by ignoring the line profile of the
emission~\citep[see][]{sara00}.  We also neglect secondary distortions
to the radiation field (but see the discussion in
Section~\ref{sec:modifications}).
These secondary distortions come from photons
emitted earlier in time, during recombination of H or He, primarily
the line transitions described in this paper.
Assuming a blackbody we have
\begin{equation}
\bar{J}(T_{\rm{M}}) = \frac{2h_{\mathrm{p}} \nu_{\alpha}^3}{c^2}
\left[ \mathrm{exp}
\left( \frac{h_{\mathrm{p}} \nu_{\alpha}}
{k_{\mathrm{B}} T_{\rm{M}}} \right) -1 \right]^{-1},
\end{equation}
where $\nu_{\alpha}= c/121.5682\,$nm$=2.466 \times 10^{15}\,$Hz and
corresponds to the energy difference between states 2p and 1s, while the
frequency of the emitted photons is equal to $\nu_{\alpha}$.
Therefore, we can set $\phi[\nu (z)] = \delta[\nu(z) -\nu_{\alpha}]$,
i.e.  a delta function centred on $\nu_{\alpha}$, so that
\begin{equation}
j_{\nu}^{\rm{Ly} \alpha}(z)=h_{\mathrm{p}} \nu \Delta
R_{2\rm{p}-1\rm{s}}^{\rm{H}}(z)
\delta[\nu(z) -\nu_{\alpha}].
\end{equation}
The increment to the intensity coming from time interval $dt$ at redshift $z$ is
\begin{equation}
d I_{\nu}(z) = \frac{c}{4 \pi} j_{\nu} dt,
\end{equation}
which redshifts to give
\[
d I_{\nu_0}(z=0) = \frac{c}{4 \pi} \frac{j_{\nu}}{(1+z)^3} dt.
\]
We assume that the emitted photons propagate freely until the present time.
Integration over frequency then gives
{\setlength\arraycolsep{2pt}
\begin{eqnarray}
I^{\mathrm{Ly} \alpha}_{\nu_0}(z=0) &=& \frac{c}{4 \pi} \int
\frac{j_{\nu}}{(1+z)^3} dt \label{inten} \\
&=& \frac{c h_{\mathrm{p}}}{4 \pi }
\frac{\Delta R_{2\rm{p}-1\rm{s}}^{\rm{H}} (z_{\alpha})}{H(z_{\alpha})
(1+z_{\alpha})^3}, \label{ILya}
\end{eqnarray}}

\noindent with
\[
1+z_{\alpha} = \frac{\nu_{\alpha}}{\nu_0},
\]
using
\[
\nu(z) = \nu_0 (1+z) \quad \mathrm{and} \quad \frac{dt}{dz}
= -\frac{1}{H(z) (1+z)}.
\]
Equation~(\ref{ILya}) is the basic equation for determining the Ly$\,\alpha$
line distortion, using $\Delta R_{2\rm{p}-1\rm{s}}^{\rm{H}}(z)$ from
the 3-level atom calculation. \\ For the two-photon emission between
the 2s and 1s levels, the emissivity at each redshift is
\begin{equation}
j_{\nu}(z)=h_{\mathrm{p}} \nu \Delta R_{2\rm{s}-1\rm{s}}^{\rm{H}}(z)
\phi[\nu(z)],
\end{equation}
and the calculation is slightly more complicated, since for
 $\phi(\nu)$ we need the frequency spectrum of the emission photons of
 the 2s--1s transition of H ~\citep{spitzer51, martinis00} as shown in
 Fig.~\ref{phi_H}.  Here $\Delta R_{2\rm{s}-1\rm{s}}^{\rm{H}}$ is the
 net rate of photon production for the 2s--1s transition, i.e.
\begin{equation}
\Delta R_{2\rm{s}-1\rm{s}}^{\rm{H}} = \Lambda_{\rm{H}}
 \left(  n_{2 \rm{s}}^{\rm{H}}
- n_1^{\rm{H}} e^{-h_{\rm{p}}\nu_{\alpha}/k_{\rm{B}}T_{\rm{M}}}\right).
\label{R2phH}
\end{equation}
Therefore, using equation~(\ref{inten}), we have
\begin{equation}
I_{\nu_0}^{2 \gamma}(z=0) = \frac{c h_p \nu_0}{4 \pi} \int_{0}^{\infty}
\frac{ \Delta R_{2\rm{s}-1\rm{s}}^{\rm{H}}  (z) \phi[\nu_0(1+z)]}{H(z)
 (1+z)^3} \ dz.
\label{dir_int}
\end{equation}
We use the simple trapezoidal rule~\citep[see Section 4.1 in][]{nr} to
integrate equation~(\ref{dir_int}) numerically from $z=0$ to the time
when $\Delta R$ is sufficiently small that the integrand can be
neglected.

%

\section{Results}
\label{result}
\begin{figure}
\centering
\vspace*{6cm} \leavevmode \includegraphics{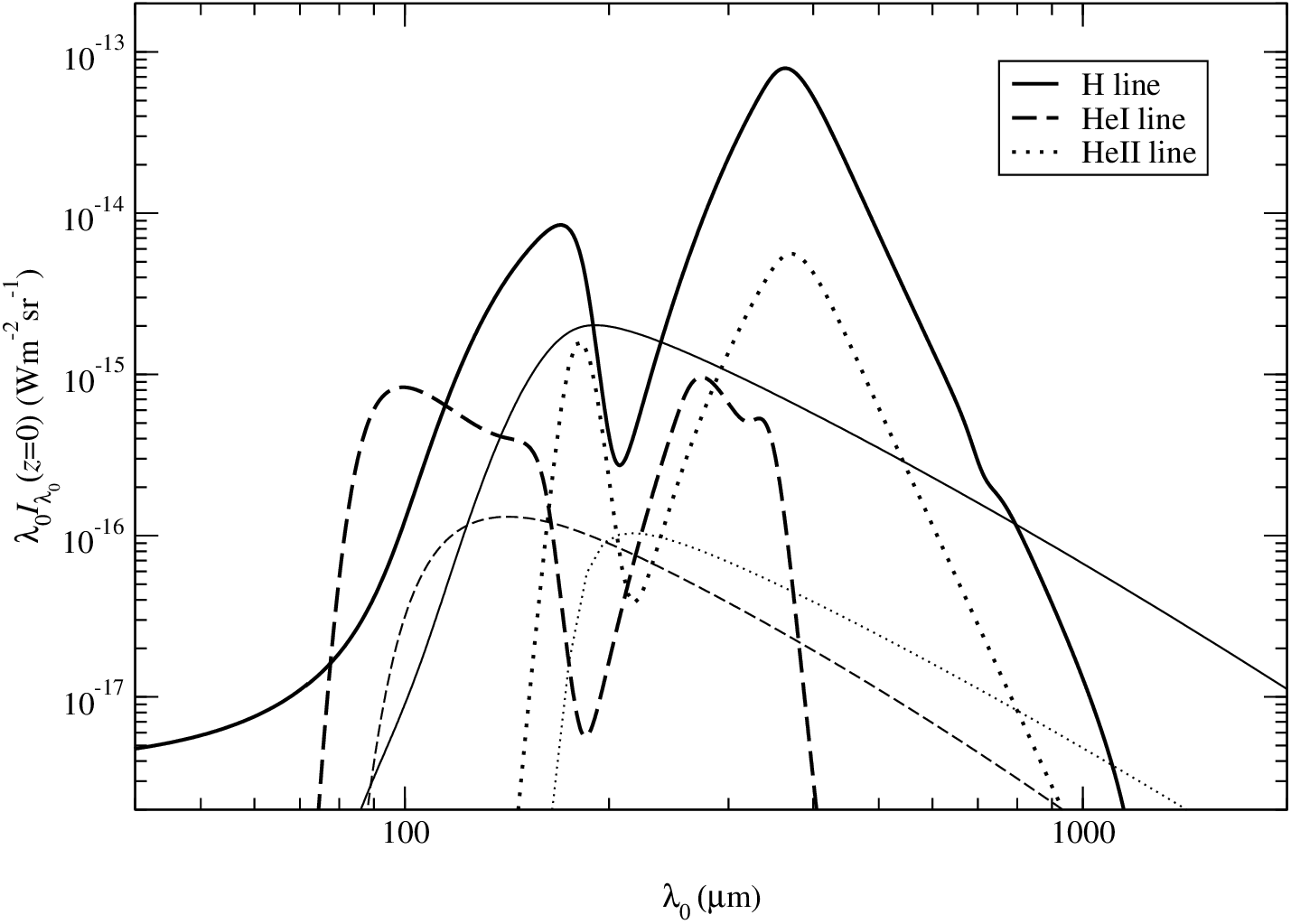}
\caption{The line intensity $\lambda_0 I_{\lambda_0}$ from the net
 Ly$\,\alpha$ emission of H (thick solid), the two-photon emission
 (2s--1s) of H with the spectrum $\phi(\nu)$ (thin solid), the
 2$^1$p--1$^1$s emission of He{\small I} (thick dashed), the
 2$^1$s--1$^1$s two-photon emission of He{\small I} (thin dashed), the
 2p--1s emission of He{\small II} (thick dotted) and the 2s--1s two-photon
 emission of He{\small II} (thin dotted).}
\label{distortHHeII}
\end{figure}

\begin{figure}
\centering
\vspace*{6cm} \leavevmode \includegraphics{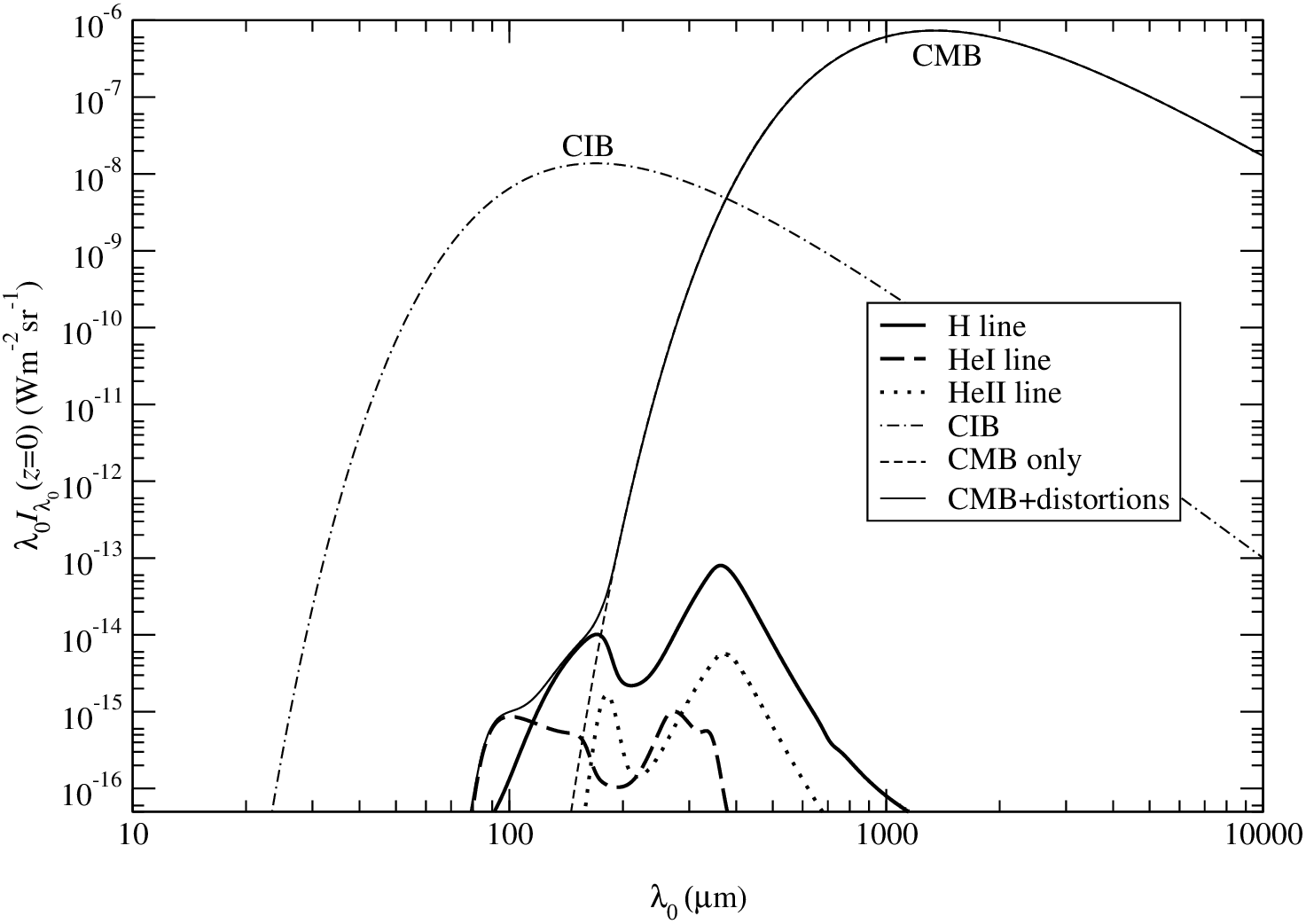}
\caption{The line intensity $\lambda_0 I_{\lambda_0}$ from the sum
of the net Ly$\,\alpha$ emission and two-photon emission (1s--2s) of H
(thick solid), the sum of the $2^1$p--$1^1$s emission and
$2^1$s--$1^1$s two-photon emission of He{\small I} (thick dashed), and
the sum of the 2p--1s emission and 2s--1s two-photon emission of
He{\small II} (thick dotted), together with the background spectra:
CMB (long-dashed); and estimated CIB ~\citep[dot-dashed;][]{fixsen98}.
The sum of all the above emission lines of H and He plus the CMB is
also shown (thin solid).}
\label{sumline}
\end{figure}

\begin{figure}
\centering
\vspace*{6cm}
\leavevmode
\includegraphics{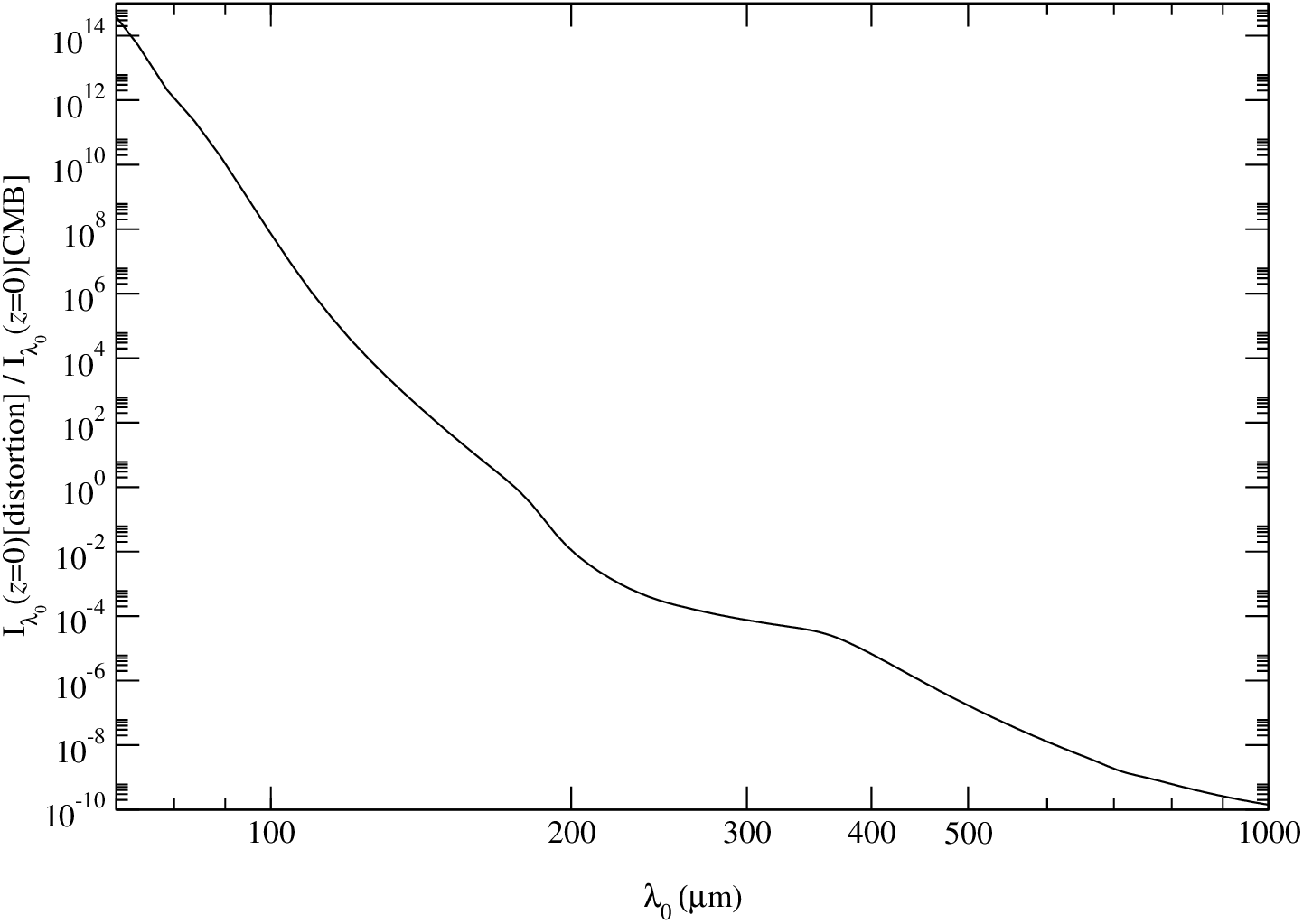}
\caption{The ratio of the total line distortion to the CMB intensity
is plotted.  The ratio is larger than 1 (i.e. the intensity of the
distortion line is larger than that of the CMB) when $\lambda_0 \sim
170\,\mu$m which is just where the main Ly$\,\alpha$ line peaks.}
\label{ratioCMB}
\end{figure}

Each of the line distortions are shown separately in
Fig.~\ref{distortHHeII} and summed for each species in
Fig.~\ref{sumline}.  The shape of the lines from H, He{\small I} and
He{\small II} are fairly similar.  There are two distinct peaks to the
2p--1s emission lines.  We refer to the one located at longer
wavelength as the `pre-recombination peak', since the corresponding
atoms had hardly started to recombine during that time.  The physics
of the formation of this peak will be discussed in detail in
section~\ref{pre_recom}.  The second (shorter wavelength)
peak is the main recombination
peak, which was formed when the atoms recombined.  While the longer
wavelength peak actually contains almost an order of magnitude more
flux, it makes a much lower relative distortion to the CMB.  The ratio
of the total distortion to the CMB intensity is shown in
Fig.~\ref{ratioCMB}.  It is approximately one for the main recombination peak,
but ${\sim}\,10^{-4}$ for the pre-recombination peak.

In Fig.~\ref{sumline}, we plot the lines from H and He{\small I}
together with the CMB and an estimate of the CIB.  We can see that the
lines which make the most significant distortion to the CMB are the
Ly$\,\alpha$ line and the 2$^1$p--1$^1$s line of He{\small I}, and
that these lines form a non-trivial shape for the overall distortion.
The sum of all the spectral lines and the CMB is shown in
Fig.~\ref{sumline}.  Note that these lines will also exist in the
presence of the CIB -- but the shape of this background is currently
quite poorly determined \citep{fixsen98, hauser98}.

We now discuss details of the physics behind the shapes of each of the
main recombination lines.

\subsection{Lines from the recombination of hydrogen}
\begin{figure}
\centering
\vspace*{6cm}
\leavevmode
\includegraphics{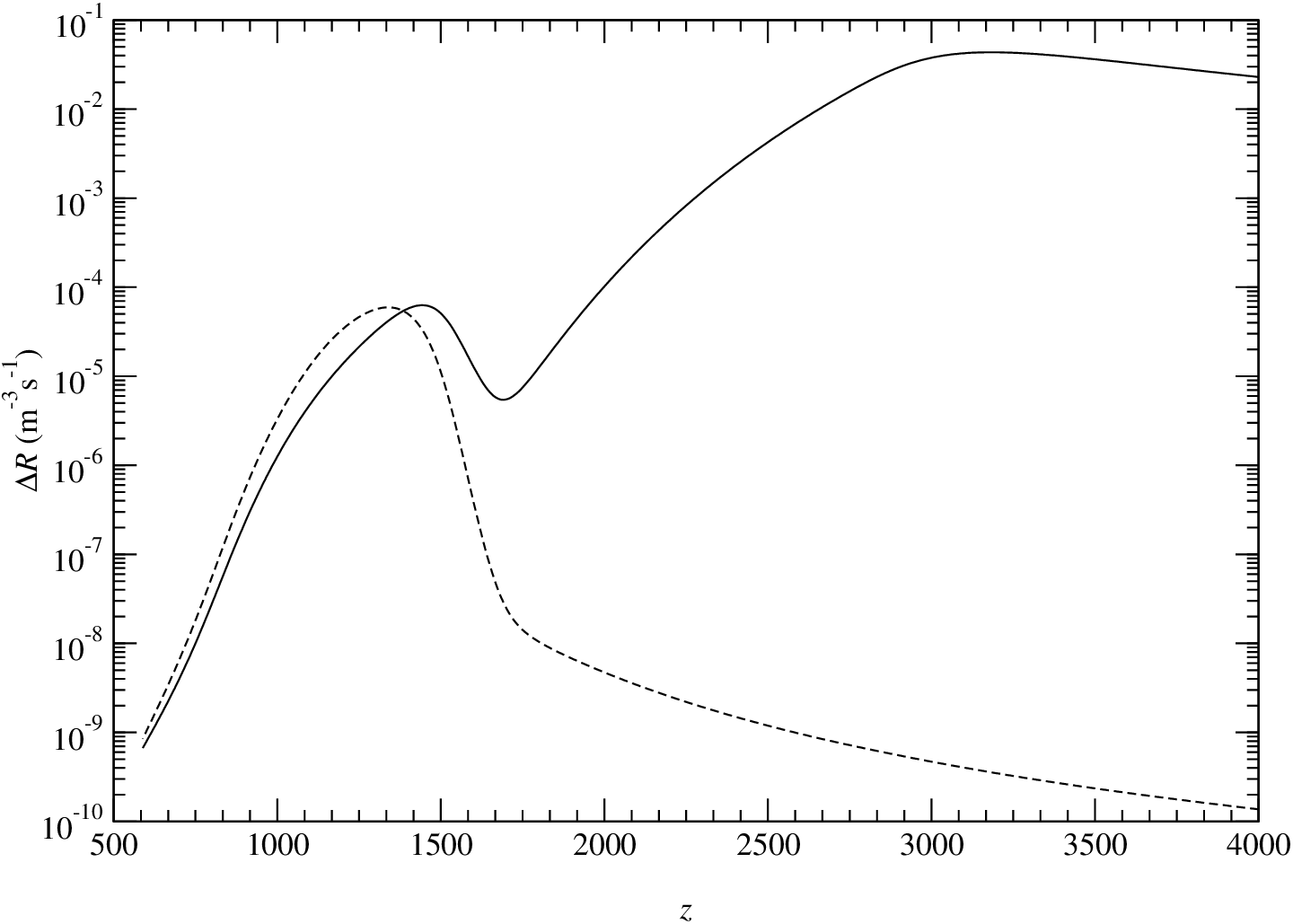}
\caption{Comparison of the net 2p--1s (solid) and 2s--1s (dashed) transition
rates of H.  The Ly$\,\alpha$ redshifting process dominates during the
start of recombination, while the 2-photon process is higher during most of the
time that recombination is occurring.  It turns out that in the standard
$\Lambda$CDM model about equal numbers of hydrogen atoms recombine through
each process, with slightly over half the hydrogen in the Universe recombining
through the 2-photon process.}
\label{ratesH}
\end{figure}

During recombination, the Lyman lines are optically thick, which means
that photons emitted from the transition to $n=1$ are instantly
reabsorbed.  However, some of the emitted photons redshift out of the line due
to the expansion of the Universe and this makes the Ly$\,\alpha$
transition one of the possible ways for electrons to cascade down to
the ground state.  The other path for electrons going from $n=2$ to
$n=1$ is the two-photon transition between 2s and 1s.  Fig.~\ref{ratesH}
shows the net photon emission rate of the Ly$\,\alpha$ and two-photon
transitions as a function of redshift for the standard $\Lambda$CDM
model.  The two-photon rate dominates at low redshift, where the bulk of
the recombinations occur.  This means that there are more photons
emitted through the two-photon emission process~(54\% of the total
number of photons created during recombination of H) than through the
Ly$\,\alpha$ redshifting process.  This conclusion agrees with
\citet{zks68} -- although of course the balance depends on the
cosmological parameters~\citep[see][]{sara00} and for today's best fit
cosmology the two processes are almost equal.  Despite this fact, the
overall strength of the two-photon emission lines are weaker because the
photons are not produced with a single frequency, but with a wide
spectrum ranging from 0 to $\nu_{\alpha}$.  The
location of the two-photon peak (see Fig.~\ref{distortHHeII}) is also
somewhat unexpected, since it is almost at the same wavelength as the
Ly$\,\alpha$ recombination peak, rather than at twice the wavelength.
The reason for this will be discussed in section~\ref{2photonpara}.

We should also note that the tiny dip in our curves for
the long-wavelength tail of the
pre-recombination peak~(see Fig.~\ref{distortHHeII}) is due to a
numerical error, when the number density of the ground state is very
small.  This can also be seen in the pre-recombination peak
for He{\small II}.

\subsubsection{The pre-recombination emission peak}
\label{pre_recom}
\begin{figure}
\centering
\vspace*{6cm} \leavevmode \includegraphics{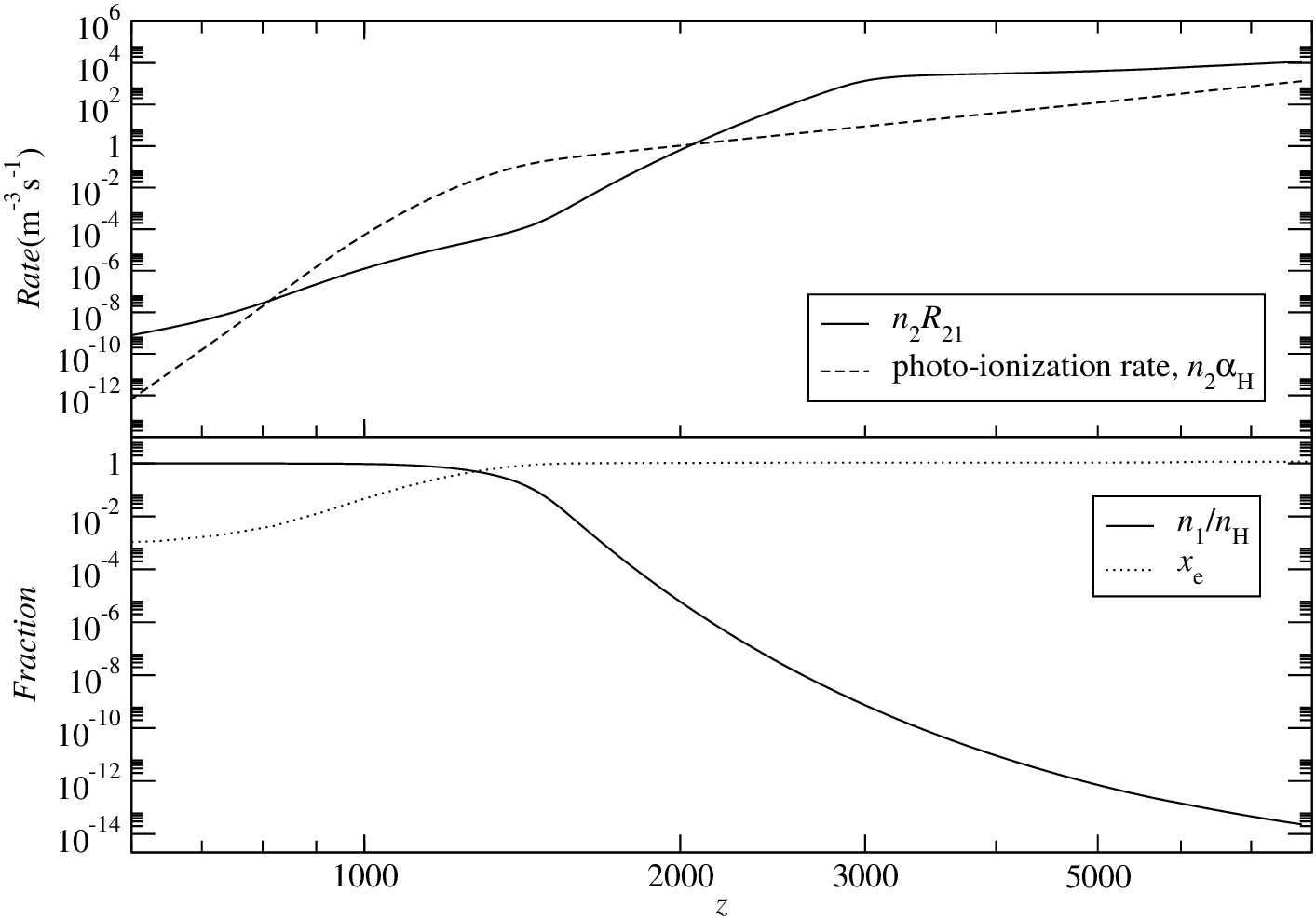}
\caption{The top panel shows the bound-bound Ly$\,\alpha$ rate $n_2
R_{21}$ and the photo-ionizing rate $n_2 \alpha_{\rm H}$ for $n$=2.
The lower panel shows the fraction of ground state H atoms $n_1/n_{\rm
H}$, and also the ionization fraction $x_{e}$.}
\label{prerec}
\end{figure}

The highest Ly$\,\alpha$ peak (shown in Fig.~\ref{distortHHeII}) is
formed before the recombination of H has already started,
approximately at $z>2000$.  During that time the emission of
Ly$\,\alpha$ photons is controlled by the bound-bound Ly$\,\alpha$
rate from $n=2$ (i.e.\ the $n_2 R_{21}$ term in equation~(\ref{RLyH}))
and the photo-ionization rate ($n_2 \alpha_{\rm H}$).  From
Fig.~\ref{prerec}, we can see that at early times the bound-bound
Ly$\,\alpha$ rate is larger than the photo-ionization rate.  This
indicates that when an electron recombines to the $n=2$ state, it is
more likely to go down to the ground state by emission of a
Ly$\,\alpha$ photon than to get ionized. 
The excess Ly alpha photons are not reabsorbed by ground state H, but
are redshifted out of the absorption frequency due to the expansion of
the Universe;  they escape freely and form the pre-recombination
 emission line.  Note that there is very little net
recombination of H, since the huge reservoir of $>13.6\,$eV CMB
photons keeps photo-ionizing the ground state H atoms~(see
Fig.~\ref{noratio}).

%
We now turn to a more detailed explanation of the pre-recombination
emission peak.  The bound-bound Ly$\,\alpha$ rate from $n=2$ is 
initially approximately constant, as
it is dominated by the spontaneous de-excitation rate (the $A_{21}$
term in equation~(\ref{R2phH})).  At the same time the
photo-ionization rate is always decreasing as redshift decreases,
since the number of high energy photons keeps decreasing with the
expansion of the Universe.  Therefore, with a constant bound-bound
Ly$\,\alpha$ rate and the decreasing photo-ionization rate, the
emission of Ly$\,\alpha$ photons rises.  The peak of this
pre-recombination line of H occurs at around $z=3000$, by which time only a
very tiny amount of ground state H atoms have formed ($n_1/n_{\rm H} <
10^{-7}$, see Fig.~\ref{prerec}).  These ground state H atoms build up
until they can reabsorb the Ly$\,\alpha$ photons and this lowers the
bound-bound Ly$\,\alpha$ rate.  The decrease of the bound-bound
Ly$\,\alpha$ rate is represented in the Sobolev escape probability
$p_{12}$ in equation~(\ref{es_prob}).  At high redshift, $p_{12}$ is 1
and there is no trapping of Ly$\,\alpha$ photons.  When H starts to
recombine, the optical depth $\tau_{\rm s}$ increases and the
Ly$\,\alpha$ photons can be reabsorbed by even very small amounts of
neutral H. For $\tau_{\rm s} \gg 1$, we can approximate $p_{12} \simeq
1/\tau_{\rm s}$ and $p_{12} \propto H(z)/n_1$.  Because of the
increase in the number density of the ground state and the decrease in
$H(z)$, the pre-recombination line decreases.  One can therefore think
of the `pre-recombination peak' as arising from direct Ly$\,\alpha$
transitions, before enough neutral H has built up to make the Universe
optically thick for Lyman photons.  This process occurs because the
spontaneous emission rate ($A_{21}$ term) is faster than the
photoionization rate for $n=2$; it increases as the Universe expands,
due to the weakening CMB blackbody radiation, and is quenched as the
fraction of atoms in the $n=1$ level grows.  The shorter wavelength
peak, on the other hand, comes from the process of redshifting out of
the Ly$\,\alpha$ line during the bulk of the recombination epoch.

By using the {\small RECFAST} program~\citep{sara99}, we can generate
the main Ly$\,\alpha$ recombination peak and also the two-photon emission
spectrum, by simply adding a few lines into the code.  However, the
pre-recombination peak cannot be generated from {\small RECFAST},
since there the rate of change of the number density of the first excited
state $n_2$ is assumed to be negligible and is related to $n_1$ via
thermal equilibrium.  Moreover, in the effective 3-level formalism,
the Ly$\,\alpha$ line is assumed to be optically thick throughout the
whole recombination process of H (in order to reduce the calculation
into a single ODE), which is not valid at the beginning of the
recombination process.  Hence, one needs to follow the rate equations
of both states (i.e. $n=1$ and $n=2$) to generate the full
Ly$\,\alpha$ emission spectrum.  The pre-recombination peak of H
was mentioned and plotted in the earlier work of \citet{dell93} as well,
although they did not describe it in any detail.

Another way to understand the line formation mechanism is to ask how
many photons are made in each process {\it per atom}.  We find that
for the main Ly$\,\alpha$ peak there are approximately 0.47 photons
per hydrogen atom (in the standard cosmology).  During the
recombination epoch, net photons for the $n=2$ to $n=1$ transitions
are only made when atoms terminate at the ground state.  Hence we
expect exactly one $n=2$ to $n=1$ photon for each atom, split between
the Ly$\,\alpha$ redshifting and 2-photon processes (and the latter
splits the energy into two photons, so there are 1.06 of these photons
per atom).  For the `pre-recombination peak', on the other hand, the
atoms give a Ly$\,\alpha$ photon when they reach $n=1$, but they then
absorb a CMB continuum photon to get back to higher $n$ or become
ionized.  The number of times an atom cycles through this process
depends on the ratio of the relevant rates.  If we take the rate per
unit volume from Fig.~\ref{ratesH} and divide by the number density of
hydrogen atoms at $z\,{\simeq}\,3000$ then we get a rate which is
about an order of magnitude larger than the Hubble parameter at that
time.  Hence we expect about 10 `pre-recombination peak' photons per
hydrogen atom.  A numerical calculation gives the more precise value
of 8.11.

\subsubsection{The two-photon emission lines}
\label{2photonpara}
Surprisingly, the location of the peak of the line intensity of the
2s--1s transition is almost the same as that of the Ly$\,\alpha$
transition, as shown in Fig.~\ref{distortHHeII}, while one might have
expected it to differ by a factor of 2.  In order to understand this
effect, we rewrite the equation~(\ref{dir_int}) in the following way:
\begin{equation}
I_{\nu_0}^{2 \gamma}(z=0) =
 \int_{0}^{\infty} \phi'(z') I^{\delta}_{\nu_0}(z=0;z') \ d z' ,
\end{equation}
where $\phi(z') = \nu_0 \phi(\nu')$, and
\begin{equation}
 I^{\delta}_{\nu_0}(z=0;z') \equiv I^{\delta}_{\nu_0}(z=0;z'(\nu')) =
 \frac{c h_{\mathrm{p}}}{4 \pi } \frac{R_{2 \gamma}(z')}{H(z')(1+z')^3}, \quad
\label{singflux}
\end{equation}
with
\[
1+z' = \frac{\nu '}{\nu_0} .
\]
Equation~(\ref{singflux}) gives the redshifted flux (measured now at
\mbox{$z=0$}) of a single frequency $\nu'$ coming from redshift $z'$
and corresponding to the redshifted frequency $\nu_0$.

\begin{figure}
\centering
\vspace*{6cm}
\leavevmode
\includegraphics{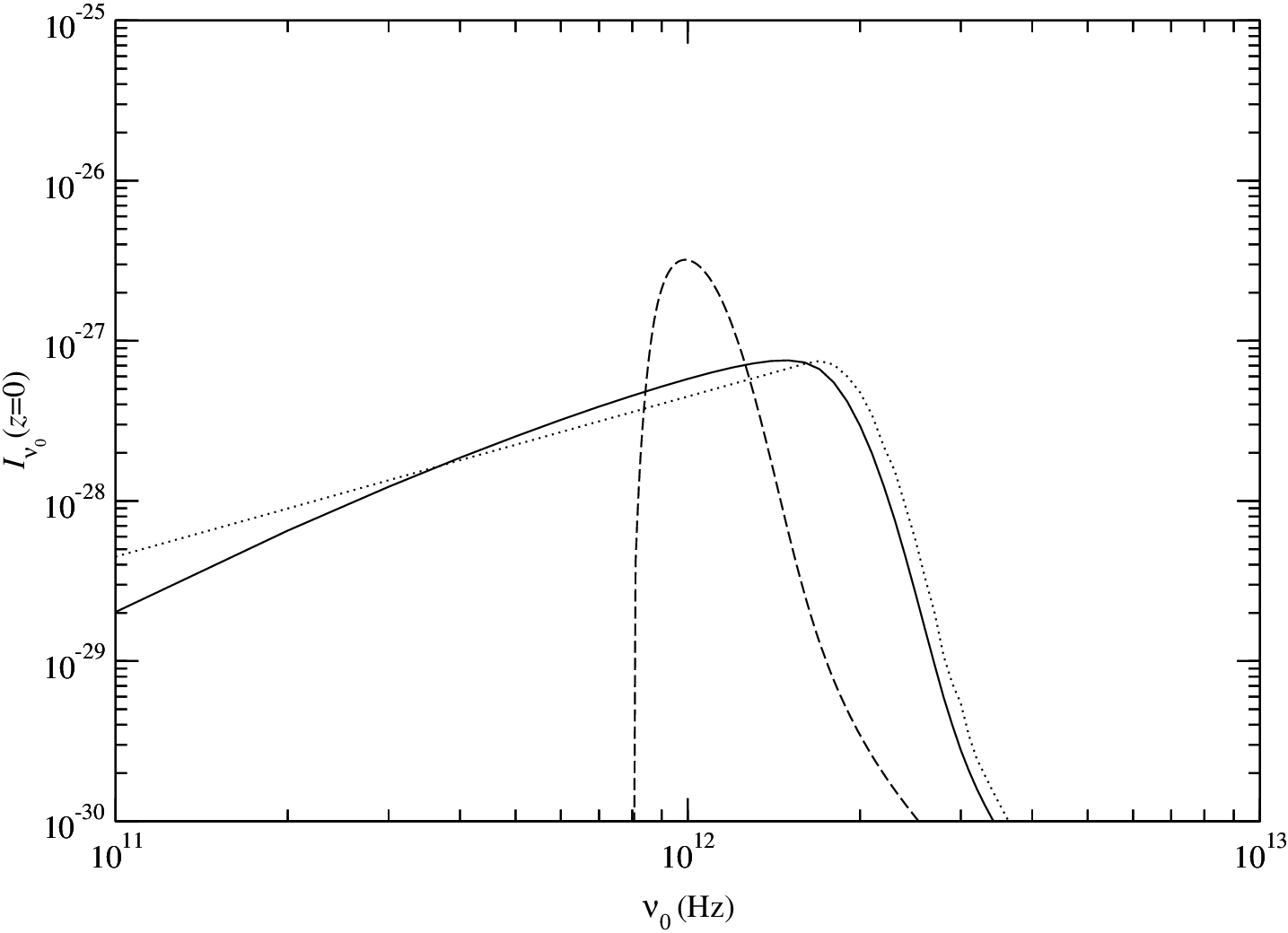}
\caption{\small The line intensity of the 2s--1s transition (two-photon
emission) $I_{\nu_0}(z=0)$ as a function of redshifted frequency
$\nu_0$ for three different assumptions: the correct frequency
spectrum of two-photon emission (solid); the delta function
approximation \mbox{$\delta(\nu - \nu_{\alpha} /2)$} (dashed); and the
flat spectrum approximation (dotted).}
 \label{Icompare}
 \end{figure}

We first calculate the line intensity of the two-photon emission with a
simple approximation: a delta function spectrum $\delta(\nu -
\nu_{\alpha}/2)$, where $\nu_{\alpha}/2$ is the frequency
corresponding to the peak of the two-photon emission spectrum
$\phi(\nu)$.  Fig.~\ref{Icompare} shows the intensity spectrum of
two-photon emission using a delta frequency spectrum $\delta (\nu-
\nu_{\alpha}/2)$ compared with the two-photon emission using the correct
spectrum $\phi(\nu)$ .  We can see that there is a significant shift
in the line centre compared with the $\delta$-function case.  Where
does this shift come from?

\begin{figure}
\centering
\vspace*{6cm}
\leavevmode
\includegraphics{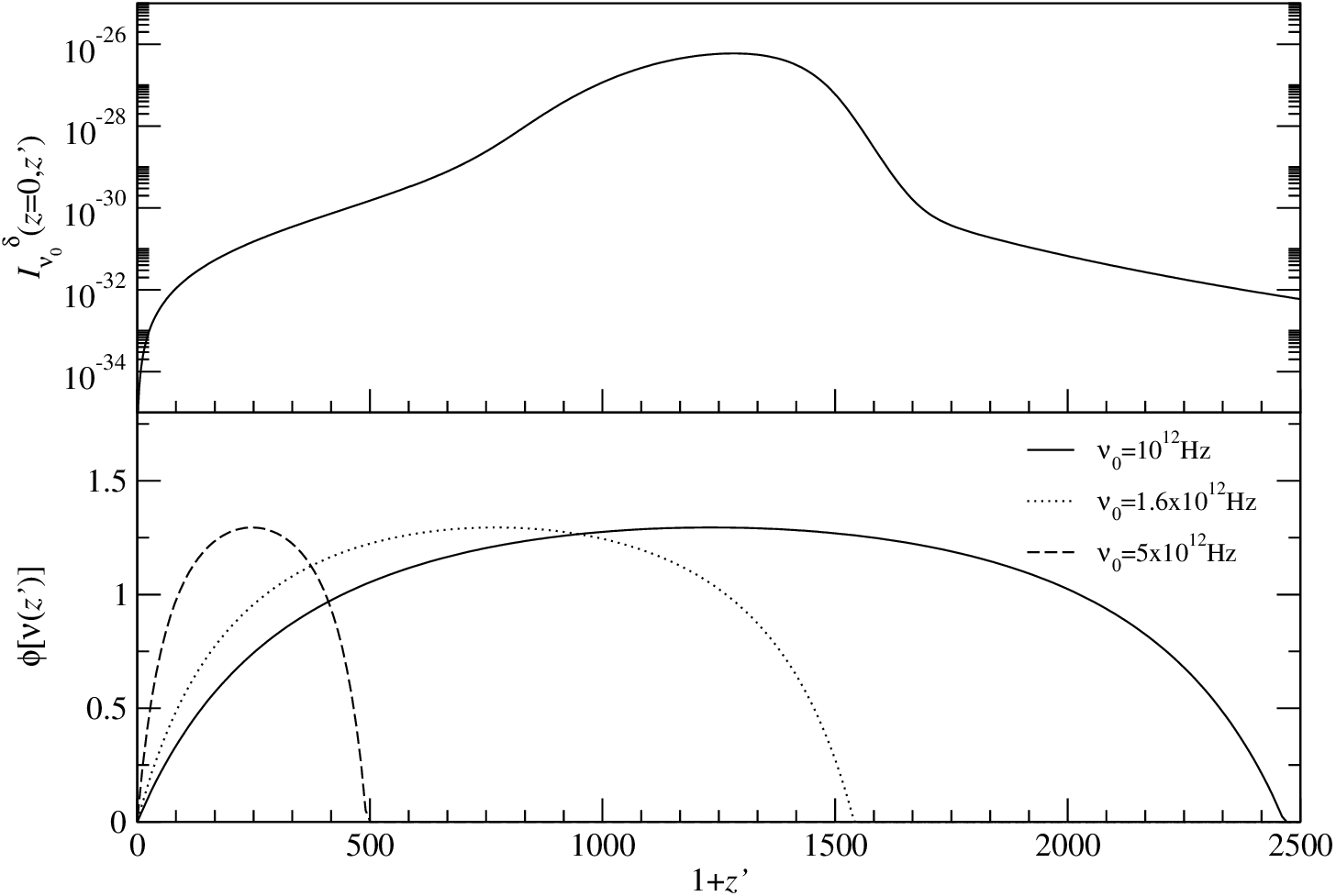}
\caption{The top panel shows the redshifted flux from single emission frequency
$I_{\nu_0}^\delta(z=0;z')$ plotted against the redshift of emission, $1+z'$.
The bottom panel shows the frequency spectrum of two-photon emission
$\phi[\nu(z')]$ plotted against $z'$ for three redshifted frequencies: $\nu_0=$
$10^{12}\,$Hz; $1.6 \times 10^{12}\,$Hz; and $5 \times 10^{12}\,$Hz.}
\label{origI2ph}
\end{figure}

We know that the frequencies of emitted photons are within the range
of $0$ to $\nu_{\alpha}$ at the time of emission.  For a fixed
redshifted frequency $\nu_0$ now, we can calculate the range of
emission redshifts contributing to $\nu_0$ (referred to as the
`contribution period' from now on), which is represented by
$\phi'(z')$ or $\phi(\nu')$ .  In Fig.~\ref{origI2ph}, we show the
spectral distribution $\phi[\nu'(z')]$ as a function of redshift $z'$
for specific values of $\nu_0$.  For example, if we take $\nu_0\,=\, 5
\times 10^{12}\,\rm{Hz}$, then photons emitted between $1+z = 1$
(i.e. $\nu=\nu_0$) and ${\sim}\,500$ ($\nu=\nu_{\alpha}$) will give
contributions to $\nu_0$.  The smaller the redshifted frequency $\nu_0$,
the wider the contribution period.  We might expect that the line intensity
of this two-photon emission will be larger if the contribution period is
longer, as there are more redshifted photons propagating from earlier
times.  However, this is not the case, because the rate of two-photon
emission $R_{2 \gamma} $ also varies with time, and is sharply peaked
at $z \simeq$ 1300--1400.  Hence $I^{\delta}_{\nu_0}(z=0;z')$ is also
sharply peaked at $z \simeq$ 1300--1400. In Fig.~\ref{origI2ph}, the
redshifted flux integrand $I_{\nu_0}^{\delta}(z=0, z)$ and the
emission spectrum $\phi[\nu(z)]$ are plotted on the same redshift
scale.  For $\nu_0= 5 \times 10^{12}\,\rm{Hz}$ (lowest panel), we can
see that the contribution period covers a redshift range when
$I_{\nu_0}^{\delta}(z=0, z)$ and $R_{2 \gamma}$ are small in value.
The contribution period widens with decreasing $\nu_0$ and covers more
of the redshift range when two-photon emission is high.  Therefore,
the flux $I_{\nu_0}(z=0)$ is expected to increase with decreasing
$\nu_0$ until the contribution period extends to the redshifts at
which the two-photon emission peaks.  As $\nu_0$ gets even smaller
(e.g.~$\nu_0=10^{12}$Hz), then the contribution period becomes larger
than the redshift range for two-photon emission and hence only lower
energy photons can be redshifted to that redshifted frequency.  As a
result, the flux $I_{\nu_0}(z=0)$ starts to decrease, and so we have a
peak. The flux peaks at $\nu_0 \simeq 10^{12}\,\rm{Hz}$ when we use
the $\delta$-function approximation.  However, from
Fig.~\ref{origI2ph}, we can see that the contribution period for
$\nu_0 \simeq 10^{12}\,\rm{Hz}$ is much greater than that of the
two-photon emission period, and therefore this is not the location of
peak.  Based on the argument presented above, we expect the peak to be
at around $1.6 \times 10^{12}\,\rm{Hz}$, or $200\,\mu$m.

The basic mathematical point is that $\phi(y)$ is extremely poorly
represented by a $\delta$-function.  Since the spectrum $\phi(\nu)$ is
quite broad, it can be better approximated as a uniform distribution
than as a $\delta$-function.  Another crude approximation
would be to assume a flat spectrum for $\phi(\nu)$ in Fig.~\ref{phi_H}.
Fig.~\ref{Icompare} compares the intensity $I_{\nu_0}(z=0)$ found
using the correct form for $\phi(\nu)$ with the $\delta$-function and
flat spectrum approximations.  This shows that the flat spectrum gives
qualitatively the same results as the correct form of the spectrum,
and that the peak occurs fairly close to that of Ly$\,\alpha$,
but is much broader.  The same general arguments apply to the
two-photon lines of He{\small I} and He{\small II} (as we discuss in
Section 3.2).

\subsubsection{Dependence of $\Omega _{\mathrm{M}}$ and $\Omega _{\mathrm{B}}$}
The largest distortion on the CMB is from the shorter wavelength
recombination peak of
the hydrogen Ly$\,\alpha$ line~(see Fig.~\ref{ratioCMB}).
It may therefore be useful to estimate the peak of this line's intensity
as a function of the cosmological parameters.  The relevant parameters
are the matter density ($\propto \Omega_{\mathrm{M}} h^2$) and the
baryon density ($\propto \Omega_{\mathrm{B}} h^2$).  This is because
$\Omega_{\mathrm{M}} h^2$ affects the expansion rate, while
$\Omega_{\mathrm{B}} h^2$ is related to the number density of
hydrogen.  No other combinations of cosmological parameters have a
significant impact on the physics of recombination.

We can crudely understand the scalings of these parameters through the
following argument.  Regardless of the escape probability $p_{12}$,
the remaining part of the rate $(n_{2 \rm{p}}^{\rm{H}} R_{21} -
n_1^{\rm{H}} R_{12})$ is roughly proportional to $n^{\rm H}_1$
$\propto \Omega_{\rm B}h^2(1-x_e)$.  The escape probability $p_{12}$
can be approximated as 1 at the beginning of recombination $(\tau_{\rm
s} \ll 1)$ and $1/\tau_{\rm s}$ during the bulk of the recombination
process (with $ \tau_{\rm s} \gg 1)$.  Note that $\tau_{\rm s} \propto
H(z)/n^{\rm H}_1 \propto (\Omega_{\rm M}h^2)^{1/2} [\Omega_{\rm
B}h^2(1-x_e) ]^{-1} $.  Therefore,
{\setlength\arraycolsep{2pt}
\begin{equation}
\Delta R_{\rm 2p-1s} \propto \left\{
\begin{array}{ll}
(\Omega_{\rm M}h^2)^{0} [\Omega_{\rm B}h^2(1-x_e) ] & {\rm for} \
\tau_{\rm s}\ll 1 \\ (\Omega_{\rm M}h^2)^{1/2} [\Omega_{\rm
B}h^2(1-x_e)]^0 & {\rm for} \ \tau_{\rm s}\gg 1,
\end{array} \right.
\end{equation}}

\noindent and thus {\setlength\arraycolsep{2pt}
\begin{equation}
I_{\lambda_0} \propto \frac{\Delta R}{H(z)} \propto \left\{
\begin{array}{ll}
(\Omega_{\rm M}h^2)^{-1/2} [\Omega_{\rm B}h^2(1-x_e) ] & {\rm for} \
\tau_{\rm s}\ll 1 \\ (\Omega_{\rm M}h^2)^{0} [\Omega_{\rm
B}h^2(1-x_e)]^0 & {\rm for} \ \tau_{\rm s}\gg 1.
\end{array} \right.
\end{equation}}

\noindent From this rough scaling argument, we may expect that the
$\Omega_{\rm M}$ dependence of the peak of the Ly$\,\alpha$ line is an
approximate power law with index between $-1/2$ and 0, while for
$\Omega_{\rm B}$ the corresponding power-law index is expected to lie
between 0 and 1.  The dependence of $\Omega_{\rm M}$ is actually more
complicated when one allows for a wider range of
values \citep[see][]{dell93}.  The above estimation just gives a rough
physical idea of the power of the dependence.

A more complete numerical estimate of the peak of the recombination
Ly$\,\alpha$ distortion is:
{\setlength\arraycolsep{0pt}
\begin{eqnarray}
 & & \left( \lambda_0 I_{\lambda_0} \right)^{\rm{peak}} \nonumber  \\
 & & \ \simeq 8.5 \times 10^{-15}
\left( \frac{\Omega_{\rm{B}} h^2 }{0.0224}  \right)^{0.57}
\left( \frac{\Omega_{\rm{M}} h^2 }{0.147}  \right)^{0.15}
\mathrm{Wm}^{-2}\mathrm{sr}^{-1},
\label{peakLya}
\end{eqnarray}}

\noindent where we have normalized to the parameters of the currently
favoured cosmological model.
The peak occurs at
\begin{equation}
\lambda_0\simeq 170\,\mu{\rm m}
\end{equation}
for all reasonable variants of the standard cosmology.

\subsection{Lines from the recombination of helium (He{\small I} and
He{\small II})}
\begin{figure}
\centering
\vspace*{6cm}
\leavevmode
\includegraphics{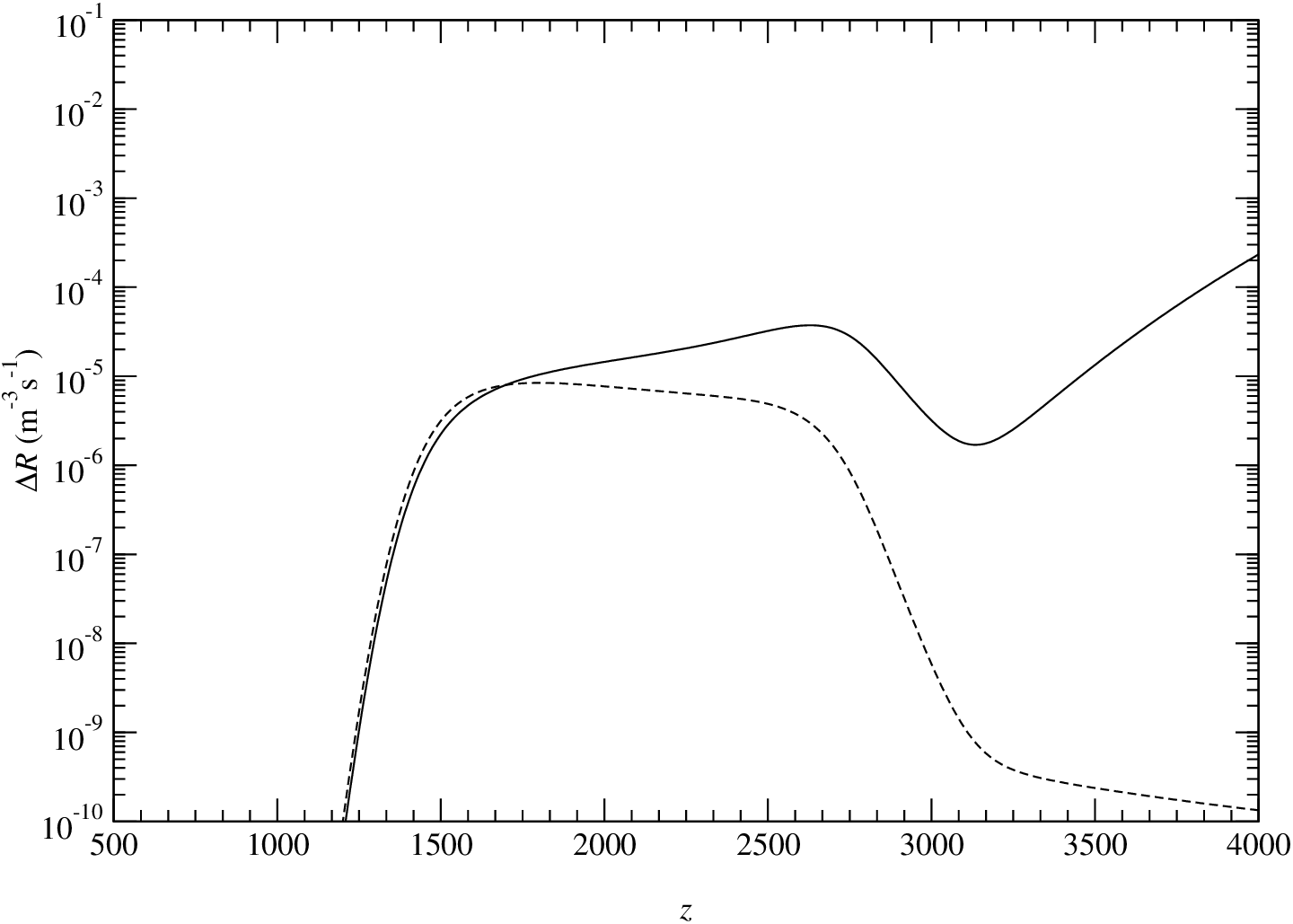}
\caption{Comparison of the net 2$^1$p--1$^1$s (solid) and
2$^1$s--1$^1$s two-photon (dashed) transition rates of He{\small I}.
The two-photon rate is sub-dominant through most of the He{\small I}
recombination epoch, and hence, unlike for hydrogen, most helium atoms did
{\it not} recombine through the two-photon process.}
\label{ratesHeI}
\end{figure}

\begin{figure}
\centering
\vspace*{6cm}
\leavevmode
\includegraphics{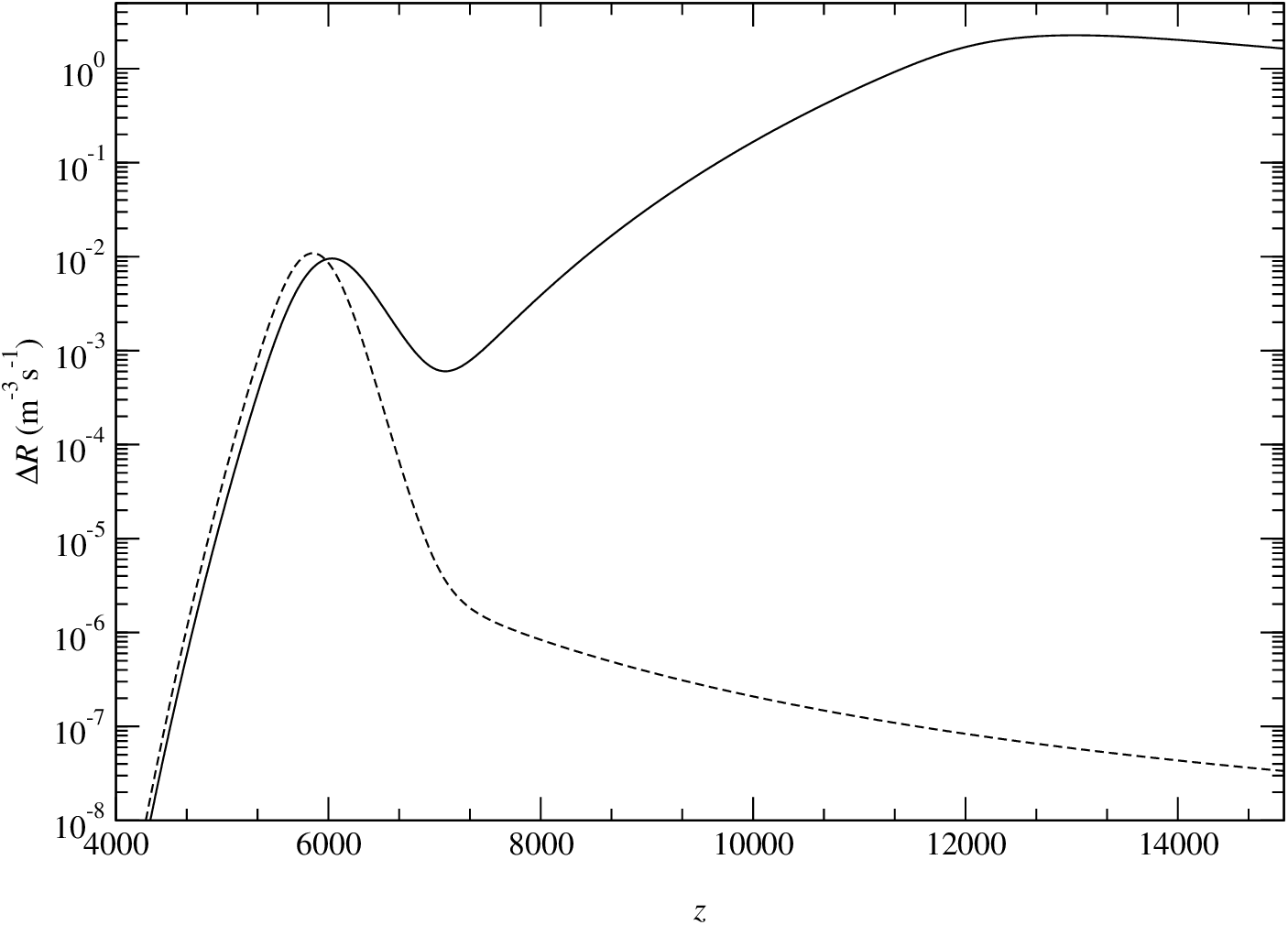}
\caption{Comparison of the net 2p--1s (solid) and
2s--1s two-photon (dashed) transition rates of He{\small II} as a function of
redshift.  The two-photon process is greater through most of the recombination
epoch, so that most of the cosmological He{\small III} $\to$ He{\small II}
process happens through the two-photon transition.}
\label{HeIIrates}
\end{figure}

We compute the recombination of He{\small II} and He{\small I} in the
same way as for hydrogen.  For the two-electron atom He{\small I}, we ignore
all the forbidden transitions between singlet and triplet states due to
the low population of the triplet states~\citep[see][]{sara99,
sara00}.  The $2^1$p--1$^1$s transitions of He{\small I} are optically thick,
the same situation as for H.  This makes the electrons take longer to
reach the ground state and causes the recombination of He{\small I} to
be slower than Saha equilibrium.  However, unlike for H, and despite
the optically thick $2^1$p--1$^1$s transition line, the $2^1$p--1$^1$s
rate dominates, as shown in Fig.~\ref{ratesHeI}.  For He{\small II},
due to the fast two-photon transition rate (see Fig.~\ref{HeIIrates}),
there is no `bottleneck' at the $n=2$ level in the recombination
process.  Hence He{\small II} recombination can be well approximated
by using the Saha equilibrium formula~\citep{sara00}.

We can see the effect of the above differences in recombination
history on the lines: the width of the recombination peak of both H
and He{\small I} is larger than that of He{\small II}.  Overall, the
spectral lines of He{\small II} are of much lower amplitude than those of H
(see Fig.~\ref{distortHHeII}) with the distortion to the CMB about an
order of magnitude smaller.

The peaks of the line distortions from H and He{\small II} are located at
nearly the same wavelengths.  For hydrogenic ions the 1s--2p energy (and all
the others) scales as $Z^2$, where $Z$ is the atomic number.  Hence for
He{\small II} recombination takes place at $z\,{\simeq}\,6000$ rather than
the $z\,{\simeq}\,1500$ for hydrogen.  Hence the line distortion from
the 2p--1s transition of He{\small II} redshifts down to about $200\,\mu$m,
just like Ly$\,\alpha$.

\begin{figure}
\centering
\vspace*{6cm}
\leavevmode
\includegraphics{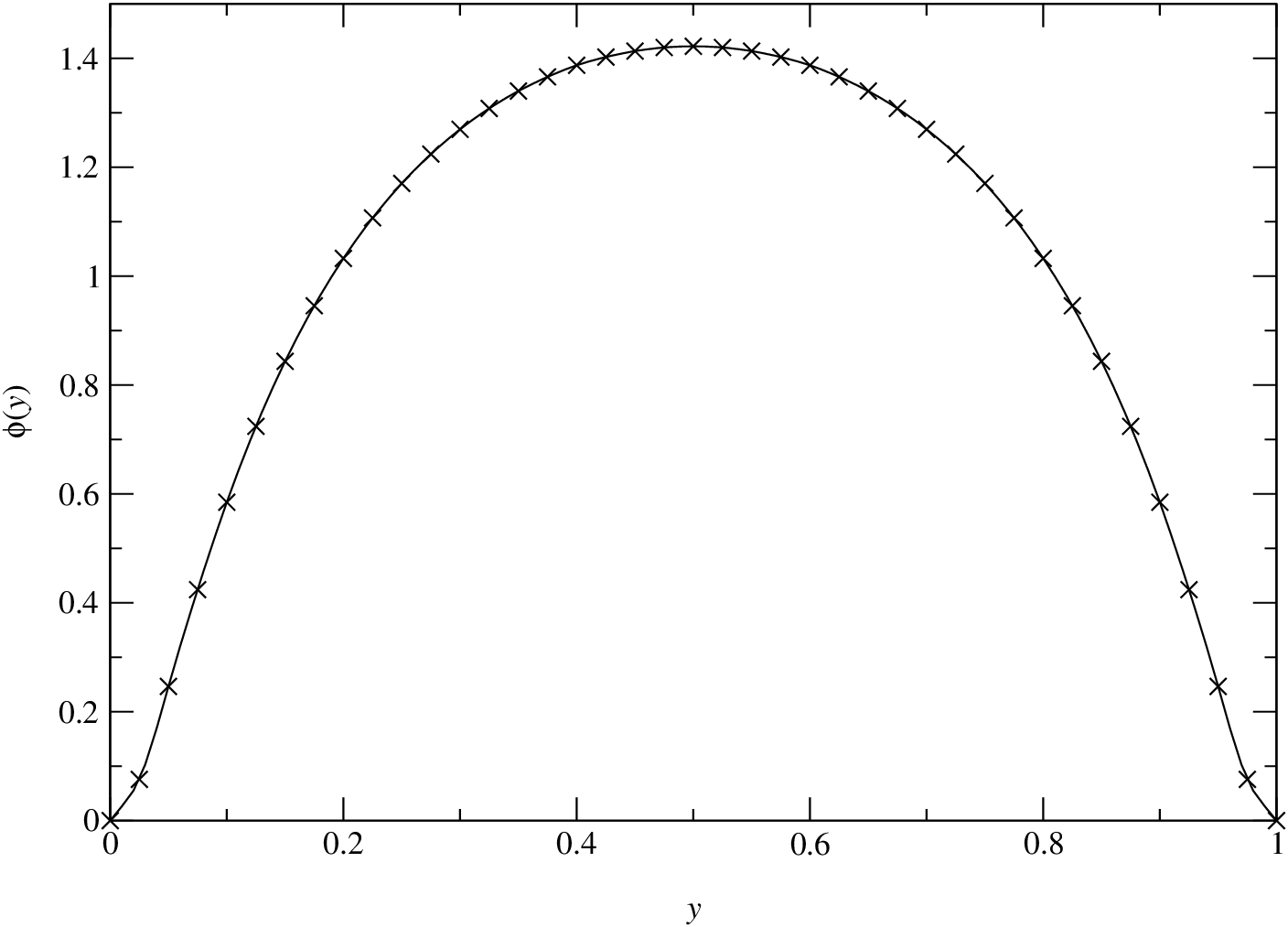}
\caption{The normalized emission spectrum for the two-photon
emission process (2$^1$s--1$^1$s) in He{\small I}.  Here $y= \nu /
\nu_{2\rm{s}-1{\rm s}}$, where $\nu_{2{\rm s} -1{{\rm s}}}=4.9849
\times 10^{15}\,$Hz.  The crosses are the calculated points from
~\citet{drake69} and \citet{drake86}, while the line is a cubic spline
fit.}
\label{phi_HeI}
\end{figure}

The two-photon frequency spectrum of He{\small II} is the same as for H, since
they both are single-electron atoms~\citep{tung84}.  However,
it is complicated to calculate the two-photon frequency spectrum of
He{\small I} very accurately, since there is no exact wave-function for the
state of the atom.  \citet{drake69} used a variational method to calculate
the two-photon frequency spectrum of He{\small I} with values given up
to 3 significant figures.
\citet{drake86} presented another calculation, giving one more
digit of precision, and making the spectrum smoother, as shown in
Fig.~\ref{phi_HeI}.  These two calculations differ by only about 1\%, which
makes negligible change to the two-photon He{\small I} spectral line.

All of the H and He lines (for $n=2$ to $n=1$) are presented in
Fig.~\ref{distortHHeII} and the sum is shown as a fractional
distortion to the CMB spectrum in Fig.~\ref{ratioCMB}.
We find that in the standard cosmological model, for He{\small I}
recombination, there are about 0.67 photons
created per helium atom in the `main' $2^1$p--$1^1$s peak, 0.70 per helium atom
in the `pre-recombination peak', and 0.66 in the two-photon process.  The
numbers for He{\small II} recombination are 0.62, 0.76 and 6.85 for these three
processes, respectively.

\section{Discussion}
\label{discuss}

\subsection{Modifications in the recombination calculation}
\label{sec:modifications}
There are several possible improvements that we could make to the line
distortion calculation.  However, as we will discuss below, we do not believe
that any of them will make a substantial difference to the amplitudes of the
lines.

\begin{figure}
\centering
\vspace*{6cm}
\leavevmode
\includegraphics{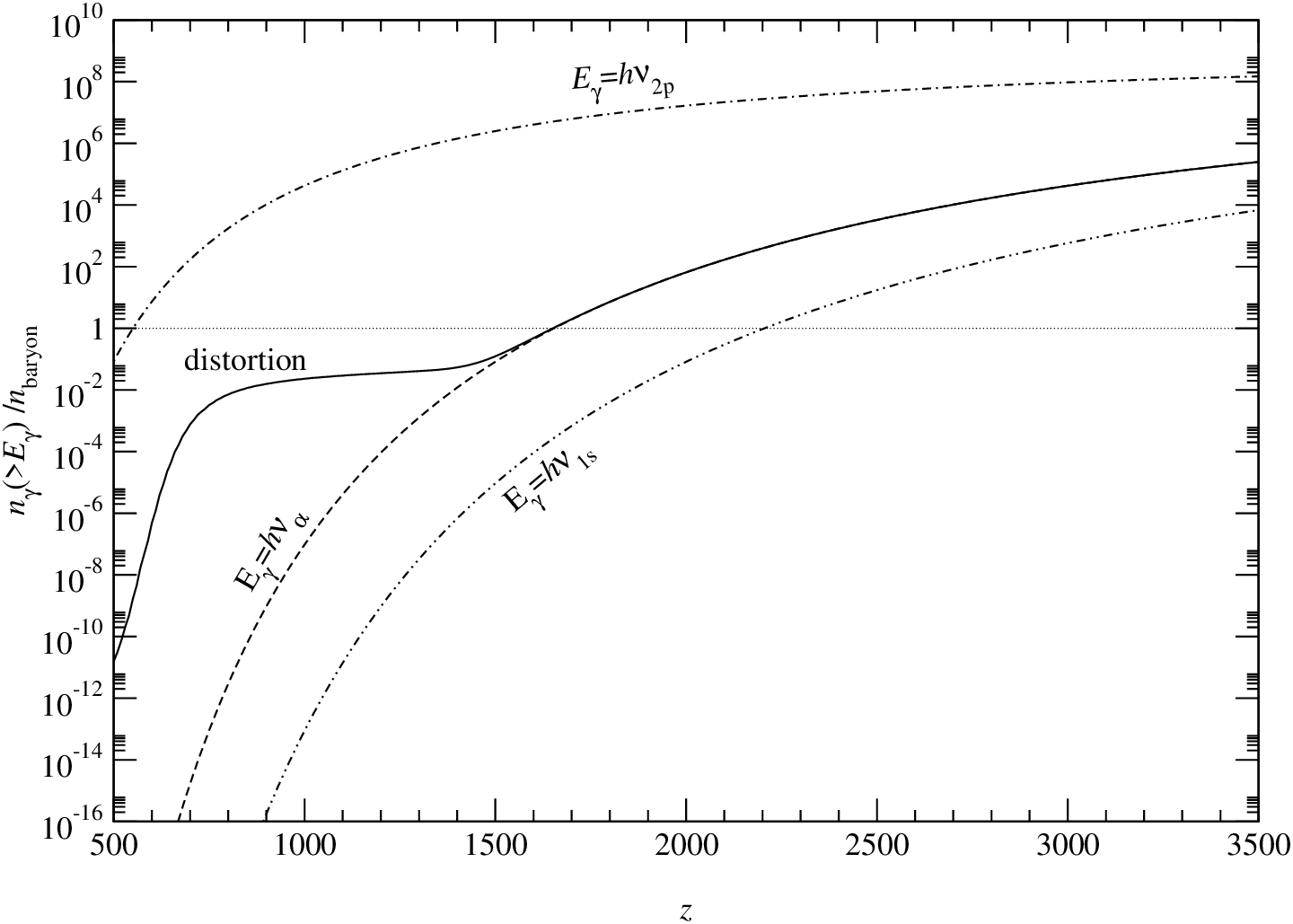}
\caption{The ratio of number of CMB photons with energy larger than
$E_{\gamma}$ ($n_{\gamma}(>E_{\gamma})$) to number of
baryons ($n_{\rm B}$) is plotted
against redshift $z$.  The solid line includes the extra distortion photons
from the recombination of He{\small I}.  From the graph, we can see that
the recombination of H occurs approximately at the redshift when the ratio of
photons with energy ${>}\,h \nu _{\alpha}$ to baryons is about unity.  By
the time the helium recombination photons are a significant distortion to
the CMB tail above Ly$\,\alpha$ the density of the relevant photons has
already fallen by 2 orders of magnitude, and so the effects can make only a
small correction.}
\label{noratio}
\end{figure}

In order to calculate the distortion lines to higher accuracy, we
should use the multi-level model without any thermal equilibrium
assumption among the bound states.  And we also need to take into
account the secondary spectral distortion in the radiation field,
i.e. we cannot approximate the background radiation field $\bar{J}$ as
a perfect blackbody spectrum.  This means, for example, that the extra
photons from the recombination of He{\small I} may redshift into an energy
range that can photoionize H($n=1$)~\citep{dell93, sara00}.
We cab assess how significant this effect might be by considering the ratio
of the number of CMB background photons with energy larger than $E_{\gamma}$,
$n_{\gamma}(> E_{\gamma})$, to the number of baryons, $n_{\rm B}$, at
different redshifts (see Fig.~\ref{noratio}).

Roughly speaking, the recombination of H occurs at the redshift when
$n_{\gamma}(> h_{\rm p} \nu_{\alpha})/n_{B}$ is about equal to 1.
This is because at lower redshifts there are not enough high energy
background photons to photo-ionize or excite electrons from the ground
state to the upper states (even to $n=2$), while at higher redshift,
when such transitions are possible, there are huge numbers of photons
able to ionize the $n=2$ level.  The solid line in Fig.~\ref{noratio}
shows the effect of the helium line distortions on the number of high
energy photons (above Ly$\,\alpha$) per baryon.  The amount of extra
distortion photons with redshifted energy larger than $h_{\rm p}
\nu_{\alpha}$ coming from the recombination of He{\small I} is only
about 1 per cent of the number of hydrogen atoms.  Their effect is
therefore expected to be negligibly small for $x_{\rm e}$.  We neglect
the effect of the helium recombination photons on the hydrogen line
distortion, since it is clearly going to make a small correction (at
much less than the 10 per cent level).

As well as this particular approximation, there have been
some other recent studies which have suggested that it may be necessary to
make minor modifications to the
recombination calculations presented in \citet{sara99, sara00}.  Although
these proposed modifications would give only small changes to the recombination
calculation, it is possible that they could have much more significant
effects on the line amplitudes and shapes.  Recent papers have described 3
separate potential effects.

In the effective three-level model,
\citet{leung04} argued that the adiabatic index of the matter should
change during the recombination process, as the ionized gas becomes
neutral, giving slight differences in the recombination history.
\citet{dub05} have claimed that the two-photon
rate between the lowest triplet state and the ground state and that between
the upper singlet states and the ground state should not be ignored in the
recombination of He{\small I}.
And \citet{chluba05} suggested that one should also include
stimulated emission from the 2s state
of H, due to the low frequency photons in the CMB blackbody spectrum.
Even if all of these effects are entirely completely
correct, we find that the change to the amplitude of the main spectral
distortion is much less than 10 per cent.  We therefore leave the detailed
discussion of these and other possible modifications to a future paper.

\subsection{Possibility of detection}
There is no avoiding the fact that detecting these CMB spectral
distortions will be difficult.  There are 3 main challenges to
overcome: (1) achieving the required raw sensitivity; (2) removing the
Galactic foreground emission; and (3) distinguishing the signal from
the CIB.

Let us start with the first point.  We can estimate the raw
sensitivity achievable in existing or planned experiments (even although
these instruments have {\it not\/} been designed for measuring the
line distortion).  Since the relevant wavelength range is essentially
impossible to observe from the ground, it will be necessary to go into space,
or at least to a balloon-based mission.  One existing experiment with
sensitivity at relevant wavelengths is BLAST (Devlin et
al.~2004), which has an array of bolometers operating at $250\,\mu$m
on a balloon payload.  The estimated sensitivity is $236\,$mJy in
$1\,$s, for a 30 arcsec FWHM beam, which corresponds to $\lambda
I_\lambda=1.2\times10^{-7}{\rm W}\,{\rm m}^{-2}{\rm sr}^{-1}$.
Comparing with equation~(\ref{peakLya}) for the peak intensity, it would take
${\sim}\,10^7$ such detectors running for a year to detect the line
distortion.  The SPIRE instrument on {\sl Herschel\/} will have a
similar bolometer array, but with better beamsize.  The estimated
sensitivity of $2.5\,$mJy at $5\sigma$ in 1 hour for a 17.4 arcsec
FWHM beamsize (Griffin, Swinyard \& Vigroux 2001) corresponds to
$\lambda I_\lambda=4.4\times10^{-8}{\rm W}\,{\rm m}^{-2}{\rm sr}^{-1}$
per detector for the $1\sigma$ sensitivity in $1\,$second.  So
detection of the line would still require ${\sim}\,10^6$ such detectors
operating for a year.

These experiments are limited by thermal emission from the instrument itself,
and so a significant advance would come from cooling the telescope.  This is
one of the main design goals of the proposed {\sl SAFIR\/} (Leisawitz 2004)
and {\sl SPICA\/} (Nakagawa et al.~2004) missions.
One can imagine improvements of a factor ${\sim}\,100$
for far-IR observations with a cooled mirror.  This would put us in the regime
where arrays of ${\sim}\,10^4$ detectors (of a size currently being
manufactured for sub-mm instruments) could achieve the desired sensitivity.

One could imagine an experiment designed to have enough spectroscopic
resolution to track the shape of the expected line distortion.  The minimum
requirement
here is rather modest, with only $\lambda/\delta\lambda\sim10$ in at least
3 bands.  An important issue will be calibration among the different
wavelengths, so that the non-thermal shape can be confidently measured.
To overcome this, one might consider the use of direct spectroscopic
techniques rather than filtered or frequency-sensitive bolometers.

Another way of quoting the required sensitivity is to say that any experiment
which measures the recombination line distortion would have to measure the
CIB spectrum with a precision of about 1 part in $10^5$, which is obviously
a significant improvement over what can be currently achieved.  A detection
of the line distortion might therefore naturally come out of an extremely
precise measurement of the CIB spectrum, which would also constrain other
high frequency distortions to the CMB spectrum.

Some of the design issues involved in such an experiment are discussed
by \citet{fixsen02}.  They describe a future experiment for measuring
deviations of the CMB spectrum from a perfect blackbody form, with an
accuracy and precision of 1 part in $10^6$.  This could provide upper
limits on Bose-Einstein distortion $\mu$ and Compton distortion $y$
parameters at the ${\sim}\,10^{-7}$ level~\citep[the current upper
limits for $y$ and $\mu$ are $15 \times 10^{-6} $ and $9 \times
10^{-5}$, respectively;][]{fixsen96}.  The wavelength coverage they
discuss is 2--120$\,{\rm cm}^{-1}$ (about 80--5{,}000$\,\mu$m), which
extends to much longer wavelengths than necessary for measuring the
line distortion.  The beam-size would be large, similar to FIRAS, but
the sensitivity achieved could easily be 100 times better.  An
experiment meant for detecting the line distortion would have to be
another couple of orders of magnitude more sensitive still.

Turning to the second of the major challenges, it will be necessary to detect
this line in the presence of the strong emission from our Galaxy.  At
$100\,\mu$m the Galactic Plane can be as bright as
${\sim}\,10^3{\rm MJy}\,{\rm sr}^{-1}$ which is about a billion times brighter
than the signal we are looking for!  Of course the brightness falls
dramatically as one moves
away from the Plane, but the only way to confidently avoid the
Galactic foreground is to measure it and remove it.  So any experiment designed
to detect the line distortion will need to cover some significant part of the
sky, so that it will be possible to extrapolate to the cosmological
background signal.
The spectrum of the foreground emission is likely to be smoother than
that of the line distortion, and it may be possible to use this fact to
effectively remove it.  However, it seems reasonable to imagine that the
most efficient separation of the signals will involve a mixture of spatial
and spectral information, as is done for CMB data (see e.g.~Patanchon
et al.~2004).

In the language of spherical harmonics, the signal we are searching for is a
monopole, with a dipole at the ${\sim}\,10^{-3}$ level and smaller angular
scale fluctuations of even lower amplitude.  Hence we would expect to be
extrapolating the Galactic foreground signals so that we can measure the
overall DC level of the sky.  This is made much more difficult by the
presence of the CIB, which is also basically a monopole signal.  Hence spatial
information cannot be used to separate the line distortion from the CIB.
The measurement of the line distortion is therefore made much more difficult
by the unfortunate fact that the CIB is several orders of magnitude brighter
-- this is the third of the challenges in measuring the recombination lines.

The shape of the CIB spectrum is currently not very well characterised.
It was detected using data from the DIRBE and FIRAS experiments on
the {\sl COBE\/} satellite.  Estimates for the background
($\lambda I_{\lambda}$) are:
$9\,{\rm nW}\,{\rm m}^{-2}{\rm sr}^{-1}$ at $60\,\mu$m
(Miville-Desch{\^e}nes, Lagache \& Puget~2002);
$23\,{\rm nW}\,{\rm m}^{-2}{\rm sr}^{-1}$ at $100\,\mu$m
(Lagache, Haffiner \& Reynolds~2000);
$15\,{\rm nW}\,{\rm m}^{-2}{\rm sr}^{-1}$ at $140\,\mu$m
(Lagache et al.~1999; Hauser et al.~1998);
and $11\,{\rm nW}\,{\rm m}^{-2}{\rm sr}^{-1}$ at $240\,\mu$m
(Lagache et al.~1999; Hauser et al.~1998).
In each case the detections are only at the 3--$5\sigma$ level, and the
precise values vary between different prescriptions for data analysis
(see also Schlegel, Finkbeiner \& Davis~1998; Finkbeiner, Davis \& Schlegel~2000;
 Hauser \& Dwek~2001; Wright~2004).
The short wavelength distortion of the CMB, interpretted as a measurement
of the CIB (Puget et al.~1996) can be fit with a modified blackbody with
temperature $18.5\,$K and emissivity index 0.64 (although there is
degeneracy between these parameters), which we plotted in Fig.~\ref{sumline}.

The CIB is thus believed to peak somewhere around $100\,\mu$m,
which is just about
where we are expecting the recombination line distortion.
The accuracy with which the CIB spectrum is known will have to
improve by about 5 orders of magnitude before the distortion will be
detectable.  Fortunately the spectral shape is expected to be significantly
narrower than that of the CIB -- the line widths are similar to the
$\delta z/z\sim0.1$ for the last scattering surface thickness, as opposed to
$\delta\lambda/\lambda\sim1$ for a modified blackbody shape (potentially
even wider than this, given that the sources of the CIB come from a
range of redshift $\Delta z\sim1$).

One issue, however, is how smooth the CIB will be at the level of detail with
which it will need to be probed.  It may be that emission lines, absorption
features, etc. could result in sufficiently narrow structure to obscure the
recombination features.  We are saved by 2 effects here: firstly the CIB
averaged over a large solid angle patch is the sum of countless galaxies,
and hence the individual spectral features will be smeared out; and
secondly, the far-IR spectral energy distributions of known galaxies do
{\it not\/} seem to contain strong features of the sort which might mimic
the recombination distortion (see e.g.~Lagache, Puget \& Dole~2005).
As we learn more about the detailed far-IR spectra of individual galaxies
we will have a better idea of whether this places a fundamental limit on
our ability to detect the recombination lines.

Overall it would appear that the line distortion should be detectable
in principle, but will be quite challenging in practice.

\section{Conclusions}  
We have studied the spectral distortion to the CMB due to the
Ly$\,\alpha$ and 2s--1s two-photon transition of H and the
corresponding lines of He{\small I} and He{\small II}.  Together these lines
give a quite non-trivial shape to the overall distortion.
The strength and shape of the line distortions are very sensitive to the
details of the recombination processes in the atoms.  Although the
amplitude of the spectral line is much smaller than the Cosmic
Infrared Background, the raw precision required is within the grasp of
current technology, and one can imagine designing an experiment to
measure the non-trivial line shape which we have calculated.  The
basic detection of the existence of this spectral distortion would
provide incontrovertible proof that the Universe was once a hot plasma
and its amplitude would give direct constraints on physics at the
recombination epoch.

\section{Acknowledgments}
This research was supported by the Natural Sciences and Engineering
Research Council of Canada.  S.S. is supported by the Carnegie
Institution of Washington.  S.S. thanks John Bahcall for stimulating
discussions.  We thank Argyro Tasitsiomi for helpying clarify Figure 1.

%

\bsp

\label{lastpage}

\end{document}